\documentclass[journal,12pt,onecolumn,draftclsnofoot]{IEEEtran}
\usepackage{graphicx}
\usepackage{amsmath}
\usepackage[noend]{algorithmic}
\usepackage{algorithm}
\usepackage{amsfonts}
\usepackage{amssymb}
\usepackage{bm}
\usepackage{cite}
\usepackage{color}
\usepackage{multirow}
\usepackage{tabularx}
\newtheorem{lem}{Lemma}
\newtheorem{prob}{Problem}
\newcolumntype{L}[1]{>{\raggedright\arraybackslash}p{#1}}
\newcolumntype{C}[1]{>{\centering\arraybackslash}p{#1}}
\newcolumntype{R}[1]{>{\raggedleft\arraybackslash}p{#1}}

\usepackage{times, epsfig}
\usepackage{subfig}
\newlength{\figwidth}
\begin{document}

\title{Throughput Maximization in Two-hop DF Multiple-Relay Network with Simultaneous Wireless Information and Power Transfer}

\author{Qi Gu, Gongpu Wang, Rongfei Fan, Ning Zhang, and Zhangdui Zhong
%\vspace{-12mm}

\thanks{
%Copyright (c) 2015 IEEE. Personal use of this material is permitted. However, permission to use this material for any other purposes must be obtained from the IEEE by sending a request to pubs-permissions@ieee.org.
%Manuscript received Feb. 25, 2017; revised Sep. 6, 2017; accepted Nov. 26, 2017. This work was supported in part by National Natural Science Foundation of China under Grant 61601025, Grant 61771054, Grant 61421001, Grant 61501028, and the Fundamental Research Funds for the Central Universities under Grant 2017YJS041. The review of this paper was coordinated by Dr. Oliver Holland.
%Corresponding author: Rongfei Fan (fanrongfei@bit.edu.cn).
Q. Gu and G. Wang are with the School of Computer and Information Technology, Beijing Jiaotong University, Beijing 100044, P. R. China (email:\{15112094, gpwang\}@bjtu.edu.cn).
R. Fan is with the School of Information and Electronics, Beijing Institute of Technology, Beijing 100081, P. R. China (email:fanrongfei@bit.edu.cn).
N. Zhang is with the Department of Computer Sciences, Texas A\&M University at Corpus Christi, TX, 78412, U.S. (ning.zhang@tamucc.edu).
Z. Zhang is with the State Key Laboratory of Rail Traffic Control and Safety, Beijing Jiaotong University, Beijing, China, 100044 (email:zhdzhong@bjtu.edu.cn).
}
}

%\author{Qi Gu, Gongpu Wang, \IEEEmembership{Member, IEEE} , Rongfei Fan, \IEEEmembership{Member, IEEE} and Zhangdui Zhong, \IEEEmembership{Senior Member, IEEE}}
%\date{}
\maketitle

%%%%%%%%%%%%%%% Article Body %%%%%%%%%%%%%%%%%%%%%%%%%%%%%%%%%%%%%%%%%

%%%%=======================================================

\begin{abstract}
This paper investigates the end-to-end throughput maximization problem for a two-hop multiple-relay network, with relays powered by simultaneous wireless information and power transfer (SWIPT) technique. Nonlinearity of energy harvester at every relay node is taken into account and two models for approximating the nonlinearity are adopted: logistic model and linear cut-off model. Decode-and-forward (DF) is implemented, and time switching (TS) mode and power splitting (PS) mode are considered. Optimization problems are formulated for TS mode and PS mode under logistic model and linear cut-off model, respectively. End-to-end throughput is aimed to be maximized by optimizing the transmit power and bandwidth on every source-relay-destination link, and PS ratio and/or TS ratio on every relay node. Although the formulated optimization problems are all non-convex. Through a series of analysis and transformation, and with the aid of bi-level optimization and monotonic optimization, etc., we find the global optimal solution of every formulated optimization problem. In some case, a simple yet optimal solution of the formulated problem is also derived. Numerical results verify the effectiveness of our proposed methods.
\end{abstract}


%\vspace{-4mm}
\section{Introduction} \label{s:intro}

Simultaneous wireless information and power transfer (SWIPT) is an emerging technical solution for energy-constrained wireless network and Internet of Things (IoT), which enables the transmitter to transmit power and information simultaneously to receiver via the radio frequency (RF) signal \cite{Wang_Ping,Weidang_Lu,Xin_Liu}.
To realize SWIPT, there are two modes: 1) Power splitting (PS) mode; 2) Time switching (TS) mode.
%; 3) Hybrid mode, which is a combination of PS mode and TS mode \cite{Wang_Ping}.
In PS mode, there is one power splitter at the receiver, which splits the received signal into two parts. One part is for energy harvesting (EH) and the other part is for information decoding (ID) \cite{Liu_2013_PS}.
In TS mode, the receiver switches between EH and ID alternatively, in which one round of EH and ID is called as one period \cite{Liu_2013_TS}.
%In hybrid mode, the receiver also operates periodically as the way in TS. Different from TS mode, within one period, the receiver in hybrid mode will first switch to EH and then switch to PS mode (this part will be also called as ID in the following for the ease of presentation) \cite{Saman_TWC_2016}.
By adjusting the PS ratio or TS ratio between EH and ID, the rate of data transmission and the rate of energy harvesting can be balanced.
This topic has been explored in lots of literatures \cite{Zhang_TC_2013, Xiangyun, Liu_2013_TS, Liu_2013_PS, Kim, Li_Shengyu, Zhang_MIMO,Chengshan_Xiao, Xiong,TVT_2016, TWC_2013,TVT_2018_Qi}.

A special utilization of SWIPT lies in relay network, in which one or more  relay nodes with no battery extracts both energy and information from the source signal through SWIPT and then forward the received signal (in amplify-and-forward (AF) mode) or decoded information (in decode-and-forward (DF) mode) to the destination node by using the harvested energy. The SWIPT-powered relay network can save relay node from additional power supply, and has attracted a lot research attentions\cite{TWC_2013, TCOM_2014,TCOM_2015_1,TVT_2016,TCOM_2015_2,JSAC_2015_Ju,SPL_2015,JSAC_2015_Xiong,Saman_Access,TVT_2016_Xiaodong,CL_2014_Xiaodong,TWC_2014_Poor}.

Two-hop or multiple-hop relay network are considered and combinations of various system configurations, e.g., PS or TS for implementing the SWIPT, DF or AF for implementing the relay, etc., are investigated in literatures.
Categorized by the research goal, two classes of literatures can be found.
The first class of literature focuses on analyzing the system performance, in terms of
ergodic capacity \cite{TWC_2013,TCOM_2014,TCOM_2015_1,TVT_2016},
effective throughput \cite{TCOM_2015_2},
or outage probability \cite{TVT_2016,TWC_2013,TCOM_2014,TCOM_2015_1,TWC_2014_Poor}.
Specifically,
\cite{TVT_2016} focuses on the DF relay network under PS mode;
\cite{TWC_2013} considers the AF relay network under PS mode and TS mode;
\cite{TCOM_2014} studies the AF and DF relay network under TS mode with full-duplex relay, which brings self-interference into the system;
\cite{TCOM_2015_1} investigates the AF relay network under PS mode with multiple-antenna relay and co-channel interference;
\cite{TCOM_2015_2} looks into the AF and DF relay network under TS mode;
\cite{TWC_2014_Poor} pays attention to the AF network under PS mode with multiple random distributed relay nodes in space and analyzed the associated performance under various relay selection strategies.
%\cite{JSAC_2015_IK} also considers the randomly distributed relay nodes in space but with battery. Different from previous works, neither PS nor TS is utilized.
It should be noticed that all the mentioned works in the first class investigate a two-hop relay network, among which \cite{TWC_2013,TCOM_2014,TCOM_2015_1,TCOM_2015_2} assume one relay while \cite{TVT_2016} and \cite{TWC_2014_Poor} assume multiple relays.

The second class of literature targets at maximizing some utility including the outage capacity \cite{JSAC_2015_Ju} or end-to-end throughput \cite{SPL_2015,JSAC_2015_Xiong,Saman_Access,TVT_2016_Xiaodong}, or minimizing some cost such as transmission time for given amount of data \cite{CL_2014_Xiaodong}, by optimizing PS ratio, TS ratio, etc.
Without specific clarification, two-hop relay network is set up in default in these literatures.
In \cite{JSAC_2015_Ju}, PS ratio and TS ratio are optimized under PS mode and TS mode in a DF relay network, respectively.
In \cite{SPL_2015}, multiple antennas are assumed at an AF relay, and PS ratio and antenna selection strategy are optimized jointly.
In \cite{JSAC_2015_Xiong}, beamforming vector and PS/TS ratio are optimized with multiple antennas implemented at source node, AF relay, and destination node.
In \cite{TVT_2016_Xiaodong}, PS ratio is optimized over multiple channels in PS mode.
%In \cite{Saman_TWC_2016}, PS ratio and TS ratio are optimized jointly in hybrid mode.
%\cite{Saman_Access} extends the work in \cite{Saman_TWC_2016} to a multi-hop DF relay network working in PS mode, TS mode, and hybrid mode.
In \cite{Saman_Access}, PS ratio and TS ratio are optimized respectively for a multi-hop DF relay network.
In \cite{CL_2014_Xiaodong}, time for energy harvesting, information decoding, and information forwarding at the relay nodes are scheduled jointly.

For all the previously surveyed works in SWIPT-powered relay network, linear model is assumed for the energy harvester at the relay node, which indicates that the output power of the energy harvesting circuit grows linearly with the power of input RF signal. However, measurements show that the practical energy harvesting circuit is subject to a non-linear model. Hence the mismatch of energy harvesting model in surveyed literatures will lead to the degradation of system performance.
In \cite{CL_2015_EB}, a nonlinear EH model based on logistic function is built, which fits the measurement data well. Some literatures related to SWIPT \cite{Xiong,TCOM_2017_EB} have also taken use of this non-linear model. For the ease of discussion, we will call this kind of model as logistic model.
In \cite{TWC_2018_Panos}, a linear cut-off model is used to approximate the nonlinear feature of energy harvester, which goes with the power of input RF signal constantly, then linearly, and at last constantly in \cite{TWC_2018_Panos}.
%and is called as constant-linear-constant in \cite{TWC_2018_Panos}.
The linear cut-off model is also shown to be a good approximation.
It should be noticed that when logistic model or linear cut-off model is adopted for a SWIPT-powered relay network, the methods in existing literatures cannot offer a solution.

In this paper, we investigate the two-hop DF relay network  with a consideration of nonlinear energy harvester under TS mode and PS mode for the first time. For the nonlinearity of energy harvester, both logistic model and linear cut-off model will be taken into account. The scenario with multiple relay nodes is considered, which is more general and beneficial since more copies of source signal can be utilized. Thus there are multiple links from source to destination through a relay node.
End-to-end throughput is targeted to be maximized by optimizing transmit power and bandwidth on every link and PS ratio or TS ratio on every relay node.
Optimization problems are formulated for TS mode and PS mode, respectively.
\begin{itemize}
  \item For TS mode under two nonlinear models of energy harvester, the associated optimization problem is non-convex. To find the global optimal solution, the original optimization problem is decomposed into two levels. In the lower level, with some further transformations and by exploring the special properties of investigated problem, closed-form optimal solution is derived. In the upper level, the associated problem is transformed to be a standard monotonic optimization problem, whose global optimal solution is achievable.
  \item For PS mode under logistic model, with some transformations, the original optimization is also transformed to be a standard monotonic optimization problem. Hence the global optimal solution is also achievable.
  \item For PS mode under linear cut-off model, the method for PS mode under logistic model also applies. However, to further save the computation complexity,  we transform the original optimization problem to be an equivalent form and then derive the semi-closed-form solution for the transformed problem, which is also global optimal.
%  \item For hybrid mode under logistic model and linear cut-off mode, we show that the associated optimization problem can be solved by combining the our solutions for TS mode and PS mode.
\end{itemize}

The rest of this paper is organized as follows. In Section \ref{s:model},  the system model is presented and the research problems are formulated.
Section \ref{s:optimal_solution_TS} and Section \ref{s:optimal_solution_PS} present the optimal solution of the formulated problem in TS mode and PS mode, respectively.
Section \ref{s:num} shows the numerical results, followed by concluding remarks in Section \ref{s:conclusion}.

\section{System Model and Problem Formulation} \label{s:model}

\begin{figure}[!hbtp]
\begin{center}
\includegraphics[angle=0,width=0.47 \textwidth]{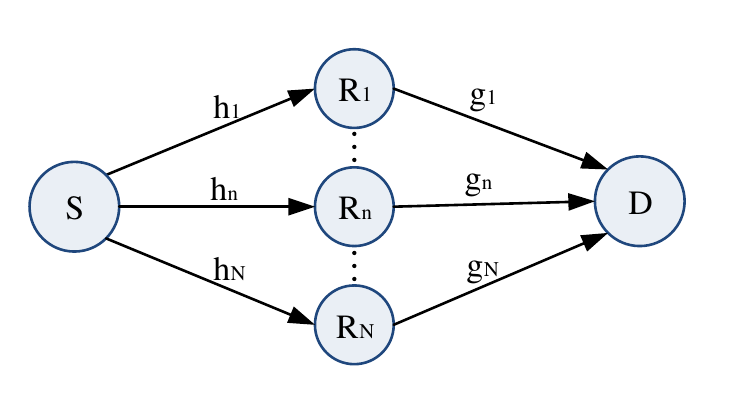}
\end{center}
\caption{Illustration of a two-hop multiple-relay network.}
\label{fig:multiple_relay_nodes}
\end{figure}

Consider a two-hop DF multiple-relay network as shown in Fig. \ref{fig:multiple_relay_nodes}, in which the source node $S$ would like to transmit information to the destination node $D$ via relay node $R_n$, who has no power supply, for $n\in \mathcal{N}  \triangleq \{1, 2, ..., N\}$.
The source, destination, and $N$ relays all have single antenna.
Denote the channel gain from $S$ to $R_n$ as $h_n$,
the channel gain from $R_n$ to $D$ as $g_n$,
and the path from node $S$ to node $D$ through node $R_n$ as link $n$, for $n \in \mathcal{N}$.
%There is also a direct link between node $S$ and node $D$, which is denoted as link 0. The channel gain of the direct link is denoted as $h_0$.
A direct link from the source node $S$ to the destination node $D$ does not exist due to physical obstacles \cite{CL_2012, TWC_2013}.
All the links also constitute the set $\mathcal{N} \triangleq \{1, 2,..., N\}$.
In the system, all the channel gains keep stable in one fading block, and are randomly and independently distributed over fading blocks with continuous distribution function.

The information is transmitted with the help of relay nodes in the following way.
%DF is adopted for every relay node.
Denote the bandwidth allocated to link $n$ as $w_n$, suppose the transmit power of source node $S$ as $p_n$ for link $n$.
By assuming the total system bandwidth as $w_T$, and total transmit power of source node $S$ as $p_T$, there are
\begin{equation}
w_n \geq 0, \forall n \in \mathcal{N},
\end{equation}
\begin{equation}
\sum \limits_{n=1}^{N} w_n \leq w_T,
\end{equation}
\begin{equation}
p_n \geq 0, \forall n \in \mathcal{N},
\end{equation}
and
\begin{equation}
\sum \limits_{n=1}^{N} p_n = p_T.
\end{equation}

For every relay node, it should be noticed that they all have no power supply.
%the constraints of every relay node's transmit power, denoted as $q_n$ for relay node $R_n$, it should be noticed that relay node $R_n$ has no power supply.
Thus $R_n$ for $n\in \mathcal{N}$ has to harvest energy from the signal transmitted by node $S$.
SWIPT technique is utilized, and two modes are considered: TS mode and PS mode.
%, and hybrid mode.
%Since the TS mode and the PS mode are special cases of the hybrid mode, we will shown the detailed procedure of hybrid mode in the following

\begin{figure}[!hbtp]
\begin{center}
\includegraphics[angle=0,width=0.47 \textwidth]{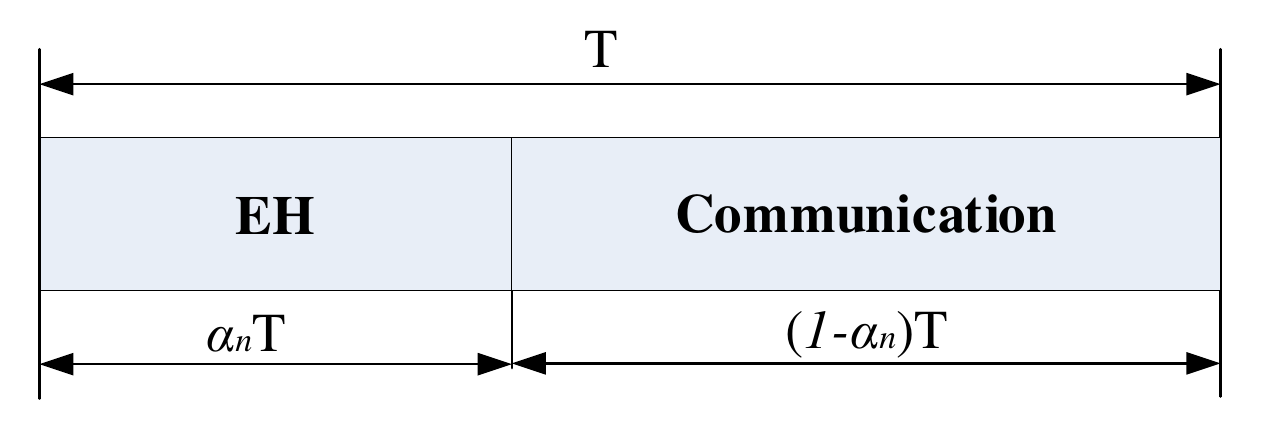}
\end{center}
\caption{Time frame structure of TS mode for relay node $R_n$.}
\label{fig:ts}
\end{figure}

In TS mode,
\begin{itemize}
\item Step 1: As shown in Fig. \ref{fig:ts}, time is divided into multiple frames with equal length $T$. The $T$ is smaller than the coherence time, hence channel gains $h_n$ and $g_n$ for $n\in \mathcal{N}$ within in $T$ keeps invariant. Within one frame, $R_n$ first harvests energy from $S$'s RF signal in the time duration between $[0, \alpha T]$, where $0 \leq \alpha \leq 1$. In this step, the harvested energy can be written as $\alpha T \phi(p_T h_n)$, where $\phi(x)$ indicates the power of harvested energy of every relay nodes's energy harvester when the power of received energy is $x$ \footnote{Without loss of generality, the feature of $\phi(x)$ of the every relay node's energy harvester is assumed to be identical.}.
\item Step 2: In the rest of time of one frame, i.e., within time duration $\left[(1-\alpha)T, T\right]$. The received signal is left for information decoding.
%Power splitting is adopted. A fraction $\beta_n$, where $0\leq \beta_n \leq 1$, of the received signal's power is left for energy harvesting, which contributes to additional harvested energy $\left(1-\alpha\right)T \phi(p_T h_n \beta_n)$ for $R_n$, and a fraction $(1-\beta_n)$ of received signal's power is left from information decoding.
\end{itemize}

\begin{figure}[!hbtp]
\begin{center}
\includegraphics[angle=0,width=0.3 \textwidth]{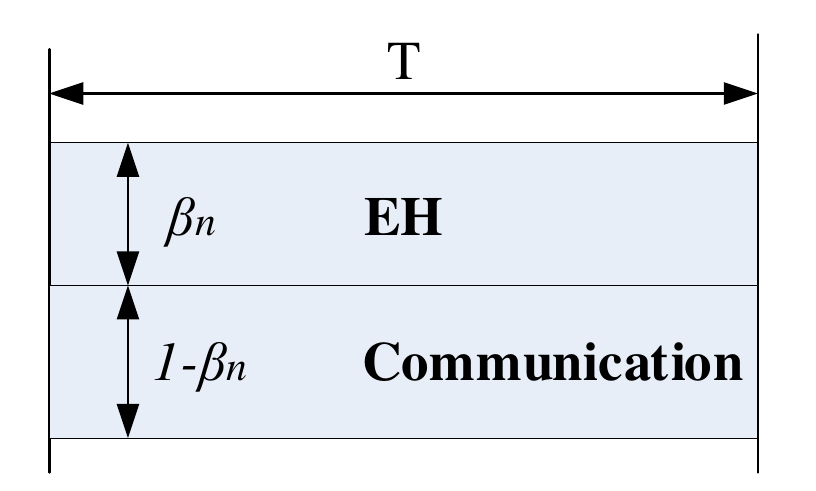}
\end{center}
\caption{Time frame structure of PS mode for relay node $R_n$.}
\label{fig:ps}
\end{figure}

In PS mode, time is also divided into multiple frames with equal length $T$, within which $h_n$ and $g_n$ for $n\in \mathcal{N}$ keeps invariant. But different from TS mode, as shown in Fig. \ref{fig:ps}, a fraction $\beta_n$ where $0\leq \beta_n \leq 1$, of the received signal's power is left for energy harvesting, and a fraction $(1-\beta_n)$ of received signal's power is left for information decoding.

%\begin{itemize}
%\item Step 1: As shown in Fig. \ref{fig:hybrid}, time is divided into multiple frames with equal length $T$. The $T$ is smaller than the coherence time, hence channel gains $h_n$ and $g_n$ for $n\in \mathcal{N}$ within in $T$ keeps invariant. Within one frame, $R_n$ first harvests energy from $S$'s RF signal in the time duration between $[0, \alpha T]$, where $0 \leq \alpha \leq 1$. In this step, the harvested energy can be written as $\alpha T \phi(p_T h_n)$, where $\phi(x)$ indicates the power of harvested energy of every relay nodes's energy harvester when the power of received energy is $x$ \footnote{Without loss of generality, the feature of $\phi(x)$ of the every relay node's energy harvester is assumed to be identical.}.
%\item Step 2: In the rest of time of one frame, i.e., within time duration $\left[(1-\alpha)T, T\right]$. Power splitting is adopted. A fraction $\beta_n$, where $0\leq \beta_n \leq 1$, of the received signal's power is left for energy harvesting, which contributes to additional harvested energy $\left(1-\alpha\right)T \phi(p_T h_n \beta_n)$ for $R_n$, and a fraction $(1-\beta_n)$ of received signal's power is left from information decoding.
%\end{itemize}
%When $\alpha=0$, the hybrid mode will turn to be PS mode.
%When $\beta_n=0$ for $n\in \mathcal{N}$, the hybrid mode dwells into TS mode.

Denote the transmit power of $R_n$ as $q_n$ for $n \in \mathcal{N}$.
In TS mode, the transmit power $q_n =  \frac{\alpha \phi(p_T h_n)}{(1-\alpha)}$.
In PS mode, the transmit power $q_n = \phi(p_T h_n \beta_n))$.
Thus the end-to-end throughput in TS mode can be written as
\begin{equation}
\begin{array}{lll}
& C_{t}(\alpha, \{p_n\}, \{w_n\}) \\
& \triangleq
\left(1 - \alpha \right) \sum \limits_{n=1}^N \min \bigg(w_n \log \left(1 + \frac{p_n h_n }{\sigma^2 w_n}\right), \\
& w_n \log \left(1 + \frac{\alpha \phi(p_T h_n)g_n}{(1-\alpha)\sigma^2 w_n}\right)\bigg).
\end{array}
\end{equation}
and the end-to-end throughput in PS mode can be written as
\begin{equation}
\begin{array}{lll}
& C_{p}(\{\beta_n\}, \{p_n\}, \{w_n\}) \\
& \triangleq
\sum \limits_{n=1}^N \min \bigg(w_n \log \left(1 + \frac{p_n h_n (1-\beta_n)}{\sigma^2 w_n}\right), \\
& w_n \log \left(1 + \frac{\phi(p_T h_n \beta_n)g_n}{\sigma^2 w_n}\right)\bigg).
\end{array}
\end{equation}
where $\sigma^2$ is the power spectrum density of noise
\footnote{When taking into security issue, a different throughput can be expressed and achieved as shown in \cite{Tu_1, Tu_2}. Due to the limit of space, we will only look into the ideal case without consideration of security in this work, which is also a general case in most of related literatures.}
.
On the other hand, $q_n$ should be also subject to a limit on the maximal transmit power, denoted as $q_{\max}$, due to the physical limit of the relay node $R_n$. Hence there is
\begin{equation}
 \frac{\alpha \phi(p_T h_n)}{(1-\alpha)} \leq q_{\max}, \forall n \in \mathcal{N}
\end{equation}
in TS mode, and
\begin{equation}
\phi(p_T h_n \beta_n) \leq q_{\max}, \forall n \in \mathcal{N}
\end{equation}
in PS mode.

\begin{figure}[!hbtp]
\begin{center}
\includegraphics[angle=0,width=0.47 \textwidth]{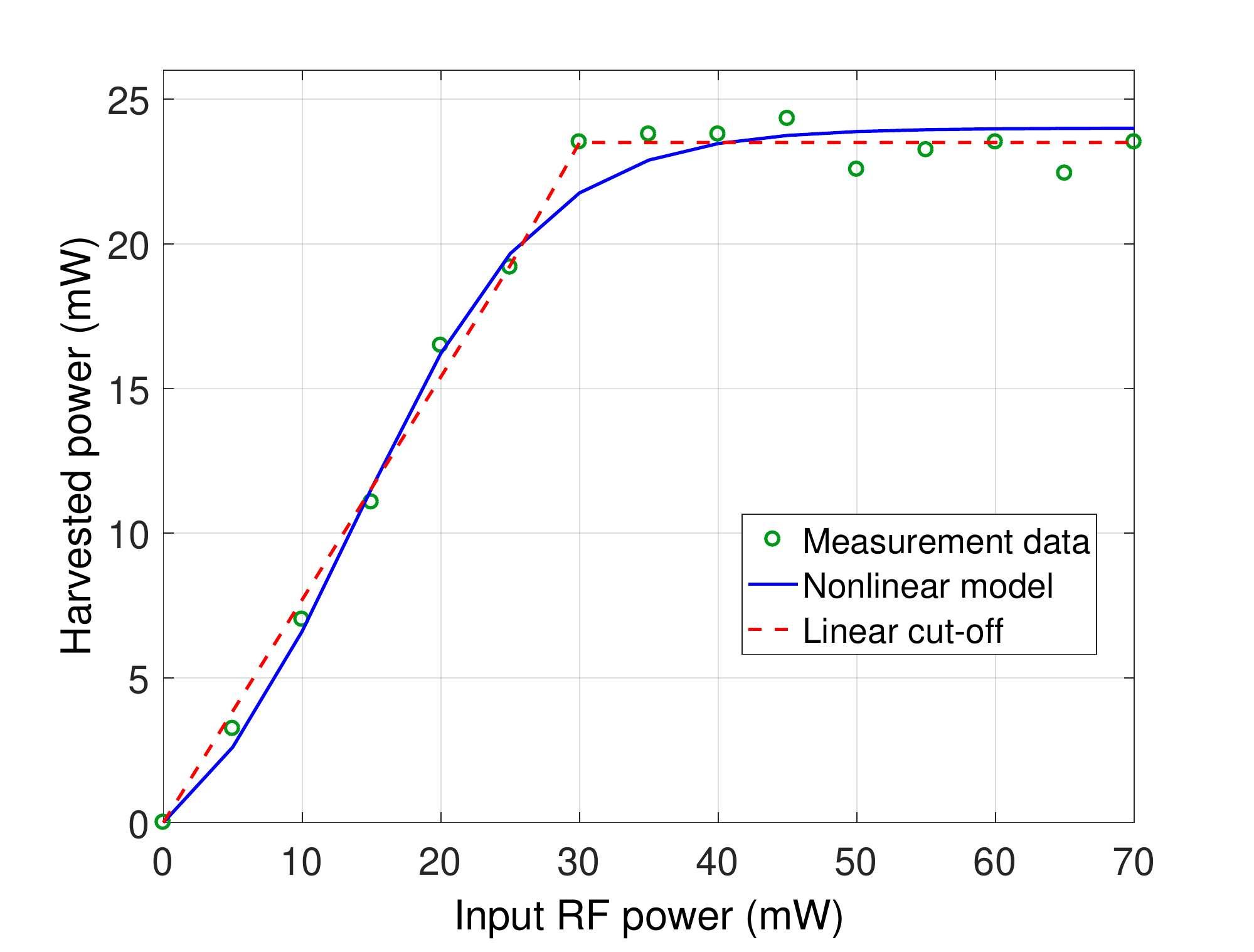}
\end{center}
\caption{Harvested power vs. input power.}
\label{fig:nonlinear}
\end{figure}

For the feature of energy harvester, as shown in Fig. \ref{fig:nonlinear}, experimental measurements in \cite{CL_2015_EB} shows that the power of harvested energy first grows with the power of received energy when the power of received energy is larger than a threshold, and then the grows slowly and slowly until it reaches up to an upper bound.
To approximate this feature, two models are adopted.
\begin{itemize}
 \item Logistic Model:
 In this model,
 \begin{equation} \label{e:model_nonlinear}
 \phi(x) = \frac{\left(\frac{M}{1+ e^{-a(x-b)}} - \frac{M}{1+e^{ab}}\right)}{\left(1 - \frac{1}{1+ e^{ab}}\right)}
 \end{equation}
 %$\phi(x) = \frac{M}{1+ e^{-a(x-b)}}$,
 where $M$ represents the maximal power the energy harvester can harvest, $a$ and $b$ are parameters for nonlinearity.
This model is broadly used when taking into account the nonlinearity of the energy harvester \cite{Xiong,TCOM_2017_EB}.

 \item Linear Cut-off Model: In this model,
 \begin{equation} \label{e:model_linear_cutoff}
 \phi(x) = \left\{
 \begin{array}{lll}
 0, \text{when~} x<x_L, \\
 c (x - x_L), \text{when~} x_L \leq x \leq x_U, \\
 c (x_U - x_L), \text{when~} x > x_U.
 \end{array}
 \right.
 \end{equation}
\end{itemize}
Note that both the function in (\ref{e:model_nonlinear}) and the function in (\ref{e:model_linear_cutoff}) are monotonic increasing functions with $x$, which is in coordination with such an intuition: More power can be harvested when more power is received.
Fig. \ref{fig:nonlinear} also plots $\phi(x)$ versus $x$ under logistic model and linear cut-off model under selected parameter setup. It can be seen that both of these two models can achieve a good approximation of measurement data.

%In hybrid mode, the end-to-end throughput can be written as
%\begin{equation}
%\begin{array}{lll}
%& C_{h}(\alpha, \{\beta_n\}, \{p_n\}, \{w_n\}) \\
%& \triangleq
%\left(1 - \alpha \right) \sum \limits_{n=1}^N \min \bigg(w_n \log \left(1 + \frac{p_n h_n (1-\beta_n)}{\sigma^2 w_n}\right), \\
%& w_n \log \left(1 + \frac{\left(\alpha \phi(p_T h_n) + (1-\alpha)\phi(p_T h_n \beta_n)\right)g_n}{(1-\alpha)\sigma^2 w_n}\right)\bigg).
%\end{array}
%\end{equation}

Collecting the formulated constraints, the associated optimization problem under TS mode and PS mode can be given as follows.
%\begin{prob} \label{prob:hybrid}
%\begin{subequations}
%\begin{align}
%\max \limits_{\alpha, \left\{\beta_n\right\}, \left\{p_n\right\}, \left\{w_n\right\}}~ &C_{h}(\alpha, \{\beta_n\}, \{p_n\}, \{w_n\})  \nonumber \\
%\text{s.t.} \quad
%& 0 \leq \alpha \leq 1, \\
%& 0 \leq \beta_n \leq 1, \forall n \in \mathcal{N}, \\
%& p_n \geq 0, \forall n \in \mathcal{N}, \\
%& w_n \geq 0, \forall n \in \mathcal{N}, \\
%& \sum \limits_{n=1}^{N} w_n \leq w_T, \\
%& \sum \limits_{n=1}^{N} p_n = p_T.
%\end{align}
%\end{subequations}
%\end{prob}

In TS mode, the associated optimization problem is
\begin{prob} \label{prob:ts}
\begin{subequations}
\begin{align}
\max \limits_{\alpha, \left\{p_n\right\}, \left\{w_n\right\}}~ &C_{t}(\alpha, \{p_n\}, \{w_n\})  \nonumber \\
\text{s.t.} \quad
& 0 \leq \alpha \leq 1, \label{e:prob_ts_alpha_interval} \\
& p_n \geq 0, \forall n \in \mathcal{N}, \\
& w_n \geq 0, \forall n \in \mathcal{N}, \\
& \frac{\alpha \phi(p_T h_n)}{(1-\alpha)} \leq q_{\max}, \forall n \in \mathcal{N}, \label{e:prob_ts_alpha_max}\\
& \sum \limits_{n=1}^{N} w_n \leq w_T, \\
& \sum \limits_{n=1}^{N} p_n = p_T.
\end{align}
\end{subequations}
\end{prob}
In PS mode, the associated optimization problem is
\begin{prob} \label{prob:ps}
\begin{subequations}
\begin{align}
\max \limits_{\left\{\beta_n\right\}, \left\{p_n\right\}, \left\{w_n\right\}}~ &C_{p}(\{\beta_n\}, \{p_n\}, \{w_n\})  \nonumber \\
\text{s.t.} \quad
& 0 \leq \beta_n \leq 1, \forall n \in \mathcal{N}, \label{e:prob_ps_beta_interval}\\
& p_n \geq 0, \forall n \in \mathcal{N}, \\
& w_n \geq 0, \forall n \in \mathcal{N}, \\
& \phi(p_T h_n \beta_n) \leq q_{\max}, \forall n \in \mathcal{N}, \label{e:prob_ps_beta_max} \\
& \sum \limits_{n=1}^{N} w_n \leq w_T, \\
& \sum \limits_{n=1}^{N} p_n = p_T.
\end{align}
\end{subequations}
\end{prob}

In the following, we will show how to solve Problem \ref{prob:ts} and Problem \ref{prob:ps} under two energy harvester models, i.e., logistic model and linear cut-off model, respectively.

%\vspace{-4mm}
\section{Optimal Solution in TS Mode} \label{s:optimal_solution_TS}

In this section, Problem \ref{prob:ts} will be solved. Note that Problem \ref{prob:ts} is a non-convex optimization problem given that the function $C_t(\alpha,\{p_n\},\{w_n\})$ is a non-concave function with the vector of $\alpha$, $\{p_n\}$, and $\{w_n\}$. Thus the global optimal solution of Problem \ref{prob:ts} is hard to achieve. In the following, we will do some transformation and simplification on Problem \ref{prob:ts}, and find the global optimal solution of Problem \ref{prob:ts}. Attention that the presented solution in this section works for both the case under logistic model and the case under cut-off model.

To solve Problem \ref{prob:ts} optimally, we decompose it into two levels \footnote{This method is referred to as {\it bi-level optimziation}.}. In the lower level, $\alpha$ is fixed, and the following optimization problem need to be solved
\begin{prob} \label{prob:ts_lower}
\begin{subequations}
\begin{align}
F(\alpha) \triangleq \max \limits_{\left\{p_n\right\}, \left\{w_n\right\}}~ & \sum \limits_{n=1}^N \min \bigg(w_n \log \left(1 + \frac{p_n h_n }{\sigma^2 w_n}\right), \nonumber \\
& w_n \log \left(1 + \frac{\alpha \phi(p_T h_n)g_n}{(1-\alpha)\sigma^2 w_n}\right)\bigg) \nonumber \\
\text{s.t.} \quad
& p_n \geq 0, \forall n \in \mathcal{N}, \\
& w_n \geq 0, \forall n \in \mathcal{N}, \\
& \sum \limits_{n=1}^{N} w_n \leq w_T, \\
& \sum \limits_{n=1}^{N} p_n = p_T \label{e:prob_ts_lower_sum_p_equal}
%& p_n \leq \frac{\alpha \phi(p_T h_n) g_n}{(1-\alpha)h_n}, \forall n \in \mathcal{N} \label{e:prob_ts_sim_p_max}.
\end{align}
\end{subequations}
\end{prob}

For the upper level,
look into the constraint (\ref{e:prob_ts_alpha_max}), which is equivalent with
\begin{equation}
\alpha \leq \frac{q_{\max}}{q_{\max} + \phi(p_T h_n)}, \forall n \in \mathcal{N}.
\end{equation}
Define $\alpha_{\max} \triangleq \min \limits_{n\in \mathcal{N}} \frac{q_{\max}}{q_{\max} + \phi(p_T h_n)}$.
Note that $\alpha_{\max} \leq 1$. Thus the constraint (\ref{e:prob_ts_alpha_max}) and constraint (\ref{e:prob_ts_alpha_interval}) can be combined to be
\begin{equation}
0 \leq \alpha \leq \alpha_{\max}.
\end{equation}
In the upper level, we need to optimize
$\alpha$ so as to solve the following optimization problem
\begin{prob} \label{prob:ts_upper}
\begin{subequations}
\begin{align}
\max \limits_{\alpha} ~ & (1-\alpha) F(\alpha) \nonumber \\
\text{s.t.} \quad
& 0 \leq \alpha \leq \alpha_{\max}.
%& p_n \leq \frac{\alpha \phi(p_T h_n) g_n}{(1-\alpha)h_n}, \forall n \in \mathcal{N} \label{e:prob_ts_sim_p_max}.
\end{align}
\end{subequations}
\end{prob}
It can be checked that Problem \ref{prob:ts} is equivalent with the upper level optimization problem, i.e., Problem \ref{prob:ts_upper}.

%\subsection{Optimal Solution for Problem \ref{prob:ts_lower}} \label{s:ts_lower}
\subsection{Optimal Solution for the Lower Level Optimization Problem} \label{s:ts_lower}
In this subsection, we will solve the lower level optimization problem, i.e., Problem \ref{prob:ts_lower}.
To simplify the solving of Problem \ref{prob:ts_lower},
we impose one additional constraint
\begin{equation} \label{e:prob_ts_sim_p_max}
p_n \leq \frac{\alpha \phi(p_T h_n) g_n}{(1-\alpha)h_n}, \forall n \in \mathcal{N},
\end{equation}
then the objective function of Problem \ref{prob:ts_lower} reduces to
\begin{equation*}
\sum \limits_{n=1}^N w_n \log \left(1 + \frac{p_n h_n }{\sigma^2 w_n}\right).
\end{equation*}
In addition, by relaxing the equality constraint (\ref{e:prob_ts_lower_sum_p_equal}) in Problem \ref{prob:ts_lower} to be an inequality, Problem \ref{prob:ts_lower} turns to be the following optimization problem
%To facilitate the solving of Problem \ref{prob:ts_lower}, we reduce the $\min(\cdot)$ function in the objective function of Problem \ref{prob:ts} by imposing the following constraint
%and focus on the following simplified optimization problem
\begin{prob} \label{prob:ts_lower_sim}
\begin{subequations}
\begin{align}
\max \limits_{\left\{p_n\right\}, \left\{w_n\right\}}~ & \sum \limits_{n=1}^N w_n \log \left(1 + \frac{p_n h_n }{\sigma^2 w_n}\right)  \nonumber \\
\text{s.t.} \quad
& p_n \geq 0, \forall n \in \mathcal{N}, \label{e:KKT_p_larger_zero}\\
& w_n \geq 0, \forall n \in \mathcal{N}, \label{e:KKT_w_larger_zero}\\
& \sum \limits_{n=1}^{N} p_n \leq p_T, \label{e:KKT_sum_p_cons}\\
& \sum \limits_{n=1}^{N} w_n \leq w_T, \label{e:KKT_sum_w_cons}\\
& p_n \leq \frac{\alpha \phi(p_T h_n) g_n}{(1-\alpha)h_n}, \forall n \in \mathcal{N} \label{e:KKT_p_upper_bound}.
\end{align}
\end{subequations}
\end{prob}
It should be noticed that maximal achievable utility of Problem \ref{prob:ts_lower_sim} equals the maximal achievable utility of Problem \ref{prob:ts_lower}. The reason is as follows:
Even the optimal solution of Problem \ref{prob:ts_lower} does not obey the constraint (\ref{e:prob_ts_sim_p_max}),
i.e., $p_n > \frac{\alpha \phi(p_T h_n) g_n}{(1-\alpha)h_n}$, the throughput on link $n$ is still $w_n \log \left(1 + \frac{\alpha \phi(p_T h_n)g_n}{(1-\alpha)\sigma^2 w_n}\right)$, which can be achieved by setting $p_n = \frac{\alpha \phi(p_T h_n) g_n}{(1-\alpha)h_n}$. In other words, in the feasible region such that constraint (\ref{e:prob_ts_sim_p_max}) holds, the maximal achievable utility of Problem \ref{prob:ts_lower} is also achievable.
To be consistent with the constraint (\ref{e:prob_ts_sim_p_max}), the equality constraint (\ref{e:prob_ts_lower_sum_p_equal}) in Problem \ref{prob:ts_lower} is relaxed to be the inequality constraint (\ref{e:KKT_sum_p_cons}), which has no influence on equality between the maximal achievable utility of Problem \ref{prob:ts_lower} and the maximal achievable utility of Problem \ref{prob:ts_lower_sim}.
Therefore solving Problem \ref{prob:ts_lower} is equivalent with solving Problem \ref{prob:ts_lower_sim}.

It should be also noticed that the solution of Problem \ref{prob:ts_lower_sim} may not serve as the optimal solution of Problem \ref{prob:ts_lower} directly, since the optimal solution of Problem \ref{prob:ts_lower_sim} may have $\sum_{n=1}^{N} p_n < p_T$. In the real application, to get the optimal solution of Problem \ref{prob:ts_lower}, we only need to find the optimal solution of Problem \ref{prob:ts_lower_sim} in the first step, and then keeps $w_n$ unchanged for $n\in \mathcal{N}$, and enlarge $p_n$ for $n\in \mathcal{N}$ calculated by solving Problem  \ref{prob:ts_lower_sim} such that
constraint (\ref{e:prob_ts_lower_sum_p_equal}) holds.

Next we turn to solve Problem \ref{prob:ts_lower_sim}. It can be checked that Problem \ref{prob:ts_lower_sim} is a convex optimization problem since the constraints of Problem \ref{prob:ts_lower_sim} are all linear and the objective function $w_n \log \left(1 + \frac{p_n h_n}{\sigma^2 w_n} \right)$ is concave with $(w_n, p_n)^T$. Although existing method can help to find the global optimal solution, in the next we will explore some special property of Problem \ref{prob:ts_lower_sim}'s optimal solution so as to simplify the solving of Problem \ref{prob:ts_lower_sim}.

It can be checked that Problem \ref{prob:ts_lower_sim} satisfies the Slater's condition. Hence the KKT condition of Problem \ref{prob:ts_lower_sim} can serve as the sufficient and necessary condition of its optimal solution \cite{Boyd}, which can be given as follows
\begin{subequations}\label{e:KKT}
\begin{align}
& \frac{w_n h_n}{p_n h_n + w_n \sigma^2 } + \eta_n - \lambda_n-\delta = 0,
\label{e:KKT_dp}  \\
& \ln\left(1+\frac{p_n h_n}{w_n \sigma^2} \right)- \frac{p_n h_n}{p_n h_n+w_n \sigma^2}+ \mu_n - \nu = 0, \label{e:KKT_dw}  \\
& \lambda_n \left(p_n - \frac{\alpha \phi(p_T h_n) g_n}{(1-\alpha)h_n} \right)=0, \forall n \in \mathcal{N}, \label{e:KKT_p_upper} \\
& \eta_n p_n = 0, \forall n \in \mathcal{N}, \label{e:KKT_p_lower}\\
& \mu_n w_n=0, \forall n \in \mathcal{N},  \label{e:KKT_w_lower}\\
& \delta \left(p_{T}-\sum\limits_{n \in \mathcal{N}} p_n \right)=0, \label{e:KKT_p_sum} \\
& \nu \left(w_T-\sum\limits_{n \in \mathcal{N}} w_n \right) =0, \label{e:KKT_w_sum}\\
& \eta_n \geq 0, \lambda_n \geq 0,  \mu_n \geq 0,  \forall n \in \mathcal{N}, \label{e:KKT_positive_lagrange_n}\\
& \nu \geq 0, \delta \geq 0, \label{e:KKT_positive_lagrange} \\
& \text{Constraints} (\ref{e:KKT_p_larger_zero}), (\ref{e:KKT_w_larger_zero}), (\ref{e:KKT_sum_p_cons}), (\ref{e:KKT_sum_w_cons}), (\ref{e:KKT_p_upper_bound}).
\end{align}
\end{subequations}
where $\eta_n$, $\mu_n$, $\delta$, $\nu$, and $\lambda_n$ are the Lagrange multipliers associated with the constraints
(\ref{e:KKT_p_larger_zero}), (\ref{e:KKT_w_larger_zero}), (\ref{e:KKT_sum_p_cons}), (\ref{e:KKT_sum_w_cons}), (\ref{e:KKT_p_upper_bound}), respectively.

Before we start the investigation on the KKT condition listed in (\ref{e:KKT}), two facts about the optimal solutions of Problem \ref{prob:ts_lower_sim} are claimed.
\begin{itemize}
\item Define $\mathcal{A} \triangleq \{n| p_n > 0, w_n >0\}$ and $\mathcal{B} \triangleq \{n | p_n=0, w_n =0\}$. Then there is $\mathcal{N} = \mathcal{A} \cup \mathcal{B}$.  This fact indicates that the case with $p_n >0, x_n=0$ (or the case $p_n=0, x_n >0$) will not happen for the optimal solution of Problem \ref{prob:ts_lower_sim}.
This is because the case with $p_n >0, x_n=0$ (or the case $p_n=0, x_n >0$) indicates a wasteful use of power resource $p_n$ (or spectrum resource $x_n$). Higher utility can be achieved by transferring the wasted resources to the other links with positive bandwidth allocation or power allocation.
\item The constraint (\ref{e:KKT_sum_w_cons}) is active, which means that $\sum_{n=1}^{N} w_n = w_T$, for the optimal solution of Problem \ref{prob:ts_lower_sim}. This is due to the fact that the objective function of Problem \ref{prob:ts_lower_sim} is an increasing function with $w_n$ for $n\in \mathcal{N}$. So it is better to increase $w_n$ for $n\in \mathcal{N}$ as much as possible.
\end{itemize}

Then we turn to investigate the KKT condition  in (\ref{e:KKT}), which can help to prove the following lemma.
\begin{lem} \label{lem:SNR_equal}
Define $\mathcal{A} \triangleq \{n| p_n > 0, w_n >0\}$, the term $\frac{p_n h_n}{w_n \sigma^2}$ equals a constant for $n\in \mathcal{A}$.
\end{lem}
\begin{IEEEproof}
For $n \in \mathcal{A}$, there is $w_n>0$, thus it can be inferred that $\mu_n=0$ from (\ref{e:KKT_w_lower}).
Define $\gamma_n = \frac{p_n h_n}{w_n \sigma^2}$, then (\ref{e:KKT_dw}) can be rewritten as
\begin{equation}
\ln (1+\gamma_n)- \frac{\gamma_n}{1+\gamma_n} =\nu, \forall n\in \mathcal{N}    \label{e:KKT_with_SNR_dw}.
\end{equation}
The function $\ln \left(1+x\right) - \frac{x}{1+x}$ is actually a strictly increasing function with $x$ for $x>0$. Hence from (\ref{e:KKT_with_SNR_dw}) it can be concluded that $\gamma_n$ for $n\in \mathcal{A}$ equals a common value, which is denoted as $\gamma$ for the ease of presentation in the following.

This completes the proof.
\end{IEEEproof}

According to the claim in Lemma \ref{lem:SNR_equal},
there is $w_n = \frac{p_n h_n}{\gamma \sigma^2}$.
Combining with the two claimed facts for the optimal solution of Problem \ref{prob:ts_lower_sim},
it can be derived that
\begin{equation} \label{e:w_T_gamma}
w_T = \sum \limits_{n=1}^{N} w_n  = \sum  \limits_{n\in \mathcal{A}} w_n  = \sum  \limits_{n\in \mathcal{A}}   \frac{p_n h_n}{\gamma \sigma^2} = \sum \limits_{n=1}^{N} \frac{p_n h_n}{\gamma \sigma^2}
\end{equation}
which further indicates that
\begin{equation}
\gamma = \sum \limits_{n=1}^{N} \frac{p_n h_n}{w_T \sigma^2}.
\end{equation}
Therefore the objective function of Problem \ref{prob:ts_lower_sim} can be rewritten as
\begin{equation}
\begin{array}{ll}
   & \sum \limits_{n=1}^N w_n \log \left(1 + \frac{p_n h_n }{\sigma^2 w_n}\right) \\
= &  \sum \limits_{n \in \mathcal{A}} w_n \log \left(1 + \frac{p_n h_n }{\sigma^2 w_n} \right) \\
= &  \sum \limits_{n \in \mathcal{A}} w_n \log \left(1 + \gamma\right) \\
= &  w_T \log \left(1 + \sum \limits_{n=1}^{N} \frac{p_n h_n}{w_T \sigma^2}\right).
\end{array}
\end{equation}
Since maximizing $w_T \log \left(1 + \sum \limits_{n=1}^{N} \frac{p_n h_n}{w_T \sigma^2}\right)$ is equivalent with maximizing $\sum_{n=1}^{N} {p_n h_n}$,  then solving Problem \ref{prob:ts_lower_sim} is equivalent with solving the following optimization problem
\begin{prob} \label{prob:ts_lower_sim_no_w}
\begin{subequations}
\begin{align}
\max \limits_{\left\{p_n\right\}}~ & \sum \limits_{n=1}^{N} {p_n h_n} \nonumber \\
\text{s.t.} \quad
& p_n \geq 0, \forall n \in \mathcal{N}, \label{e:KKT_p_larger_zero_v2}\\
& \sum \limits_{n=1}^{N} p_n \leq p_T, \label{e:KKT_sum_p_cons_v2}\\
& p_n \leq \frac{\alpha \phi(p_T h_n) g_n}{(1-\alpha)h_n}, \forall n \in \mathcal{N} \label{e:KKT_p_upper_bound_v2}.
\end{align}
\end{subequations}
\end{prob}

For Problem \ref{prob:ts_lower_sim_no_w}, it is straightforward to see that the optimal policy is to allocate more power resource to the link with higher channel gain, i.e., to set the $p_n$ with higher $h_n$ as large as possible. Specifically, the optimal allocation of $p_n$ for $n\in \mathcal{N}$ can be found as follows.
\begin{algorithm}[H]
\caption{Searching procedure for the optimal solution of Problem \ref{prob:ts_lower_sim_no_w}.}
\label{alg:p_descending_search}
\begin{algorithmic}[1]
 \STATE {Order $h_n$ for $n\in \mathcal{N}$ in descending order, such that $h_{s_1} \geq h_{s_2} \geq ... \geq h_{s_N}$.}
 \STATE {Define $i^* = \mathop{\arg \min} \limits_{i} \sum_{j=1}^{i} \frac{\alpha \phi(p_T h_{s_{j}}) g_{s_{j}} } { (1-\alpha)h_{s_{j}} } > p_T$.  Set $p_{s_i} = \frac{\alpha \phi(p_T h_{s_{i}}) g_{s_{i}} } { (1-\alpha)h_{s_{i}} }$ for $i=1, 2, ..., (i^*-1)$, $p_{s_{i}} = p_T - \sum_{j=1}^{i^*-1} \frac{\alpha \phi(p_T h_{s_{j}}) g_{s_{j}} } { (1-\alpha)h_{s_{j}} }$ for $i=i^*$, and $p_{s_{i}} = 0$ for $i=i^*+1$, $i=i^*+2$, ..., $i=N$. Note that when $\sum_{j=1}^{N} \frac{\alpha \phi(p_T h_{s_{j}}) g_{s_{j}} } { (1-\alpha)h_{s_{j}} } \leq p_T$, $i^*$ does not exist and the optimal solution is $p_n = \frac{\alpha \phi(p_T h_n) g_n}{(1-\alpha)h_n}$, $\forall n \in \mathcal{N}$.
 }
 \end{algorithmic}
\end{algorithm}

In the end of this subsection, the optimal solution of the lower level optimization problem, i.e., Problem \ref{prob:ts_lower}, can be summarized as follows.
\begin{algorithm}[H]
\caption{Searching procedure for the optimal solution of Problem \ref{prob:ts_lower}.}
\label{alg:lower_level_problem}
\begin{algorithmic}[1]
 \STATE {
 By following Algorithm \ref{alg:p_descending_search}, find the optimal $p_n$ for $n\in \mathcal{N}$ of Problem \ref{prob:ts_lower_sim_no_w}.
 }
 \STATE {
 Set $w_n = \frac{w_T p_n h_n}{\sum_{n=1}^{N} p_n h_n}$ where $p_n$ is calculated in Step 1 of Algorithm \ref{alg:lower_level_problem} for $n\in \mathcal{N}$.
  }
 \STATE{
 Increase $p_n$ calculated in Step 1 to be $p'_{n}$ for $n\in \mathcal{N}$  such that $\sum_{n=1}^{N} p'_n = p_T$.
 }
 \STATE{
 Output $p'_n$ and $w_n$ for $n \in \mathcal{N}$.
 }
 \end{algorithmic}
\end{algorithm}

\subsection{Optimal Solution for the Upper Level Optimization Problem} \label{s:ts_upper}
In this subsection, we will solve the upper level optimization problem, i.e., Problem \ref{prob:ts_upper}. In the first step, there is such a lemma.
\begin{lem} \label{lem:F_mono}
The function $F(\alpha)$, which is defined in Problem \ref{prob:ts_lower}, is monotonically increasing with $\alpha$.
\end{lem}
\begin{IEEEproof}
Suppose there is $0 \leq \alpha^{\dag} \leq \alpha^{\ddag} \leq 1$.
Define the optimal solution of Problem \ref{prob:ts_lower} associated with $\alpha^{\dag}$ and $\alpha^{\ddag}$ are $p^{\dag}_n$ and $w^{\dag}_n$, and $p^{\ddag}_n$ and $w^{\ddag}_n$, respectively, for $n\in \mathcal{N}$.
Then there is
\begin{equation}
\begin{array}{ll}
F(\alpha^{\dag})  & = \sum \limits_{n=1}^N \min \bigg(w_n^{\dag} \log \left(1 + \frac{p_n^{\dag} h_n }{\sigma^2 w_n^{\dag}}\right), \nonumber \\
&\quad w_n^{\dag} \log \left(1 + \frac{\alpha^{\dag} \phi(p_T h_n)g_n}{(1-\alpha^{\dag})\sigma^2 w_n^{\dag}}\right)\bigg) \nonumber \\
& \overset{(a)}{\leq}  \sum \limits_{n=1}^N \min \bigg(w_n^{\dag} \log \left(1 + \frac{p_n^{\dag} h_n }{\sigma^2 w_n^{\dag}}\right), \nonumber \\
& \quad w_n^{\dag} \log \left(1 + \frac{\alpha^{\ddag} \phi(p_T h_n)g_n}{(1-\alpha^{\ddag})\sigma^2 w_n^{\dag}}\right)\bigg) \nonumber \\
& \overset{(b)}{\leq}  \sum \limits_{n=1}^N \min \bigg(w_n^{\ddag} \log \left(1 + \frac{p_n^{\ddag} h_n }{\sigma^2 w_n^{\ddag}}\right), \nonumber \\
& \quad w_n^{\ddag} \log \left(1 + \frac{\alpha^{\ddag} \phi(p_T h_n)g_n}{(1-\alpha^{\ddag})\sigma^2 w_n^{\ddag}}\right)\bigg) \nonumber \\
& =  F(\alpha^{\ddag})
\end{array}
\end{equation}
where $(a)$ holds is due the fact that the coefficient $\frac{\alpha^{\dag}}{1-\alpha^{\dag}} \leq  \frac{\alpha^{\ddag}}{1- \alpha^{\ddag}}$ for $\alpha^{\dag} \leq \alpha^{\ddag}$, and $(b)$ holds since the set of $p_n^{\ddag}$ and  $w_n^{\ddag}$ for $n\in \mathcal{N}$ is the optimal solution of Problem \ref{prob:ts_lower} when $\alpha=\alpha^{\ddag}$.

This completes the proof.
\end{IEEEproof}

With Lemma \ref{lem:F_mono}, the objective function of Problem \ref{prob:ts_upper} is actually the difference between two monotonically increasing function with $\alpha$, i.e., the difference between $F(\alpha)$ and $\alpha F(\alpha)$.
Thus solving Problem  \ref{prob:ts_upper} is equivalent with solving the following optimization problem
\begin{prob} \label{prob:ts_upper_differential}
\begin{subequations}
\begin{align}
\max \limits_{\alpha, z} ~ & F(\alpha) + z \nonumber \\
%(1-\alpha) F(\alpha) \nonumber \\
\text{s.t.} \quad & \alpha F(\alpha) +  z \leq  F(\alpha_{\max}),  \label{e:prob_ts_upper_differ_sum_bound}\\
& 0 \leq \alpha \leq \alpha_{\max}.
%& p_n \leq \frac{\alpha \phi(p_T h_n) g_n}{(1-\alpha)h_n}, \forall n \in \mathcal{N} \label{e:prob_ts_sim_p_max}.
\end{align}
\end{subequations}
\end{prob}
For Problem \ref{prob:ts_upper_differential}, since both $F(\alpha)$ in its objective function and $z$ in its objective function are increasing functions with $\alpha$ and $z$ respectively, the maximum of Problem \ref{prob:ts_upper_differential} can be achieved by increasing both $\alpha$ and $z$ as large as possible.
Looking into the constraint (\ref{e:prob_ts_upper_differ_sum_bound}), both $\alpha F(\alpha)$ and $z$ are increasing functions with $\alpha$ and $z$ respectively, thus the maximum of Problem   \ref{prob:ts_upper_differential} will be achieved when both $\alpha$ and $z$ reach their maximal allowable value in the feasible region of Problem \ref{prob:ts_upper_differential}, in which case there is
\begin{equation}
\alpha F(\alpha) + z = F(\alpha_{\max}),
\end{equation}
which indicates that
\begin{equation} \label{e:z_expression}
z= F(\alpha_{\max}) - \alpha F(\alpha).
\end{equation}
Replace $z$ with the expression in (\ref{e:z_expression}), the objective function of Problem \ref{prob:ts_upper_differential} turns to be $F(\alpha) - \alpha F(\alpha) + F(\alpha_{\max})$.
Since maximizing $F(\alpha) - \alpha F(\alpha) + F(\alpha_{\max})$ is equivalent with maximizing $F(\alpha) - \alpha F(\alpha) $, solving Problem \ref{prob:ts_upper_differential} is equivalent with solving Problem \ref{prob:ts_upper}.

Then we focus on solving Problem \ref{prob:ts_upper_differential}. Although being non-convex, Problem \ref{prob:ts_upper_differential} actually falls into the standard form of {\em Monotonic Optimization Problem}, whose standard form can be given as follows.
\begin{prob} \label{prob:standard_mono}
\begin{subequations}
\begin{align}
\max \limits_{\bm{x}} ~ & f(\bm{x}) \nonumber \\
\text{s.t.} \quad & g(\bm{x}) \leq 0, \\
& \bm{x}_L \leq \bm{x} \leq \bm{x}_U,
%& p_n \leq \frac{\alpha \phi(p_T h_n) g_n}{(1-\alpha)h_n}, \forall n \in \mathcal{N} \label{e:prob_ts_sim_p_max}.
\end{align}
\end{subequations}
\end{prob}
where the variable $\bm{x}$ is a multiple dimensional vector, $\bm{x}_L$ and $\bm{x}_U$ represent the lower bound and upper bound of $\bm{x}$ respectively, and both $f(\bm{x})$ and $g(\bm{x})$ are monotonically increasing functions with $\bm{x}$.
For a standard monotonic optimization problem, there is a polyblock algorithm to achieve the $\varepsilon$-optimal solution of a standard monotonic optimization problem, where $\varepsilon$ indicates the gap between the achieved utility and the global optimal utility is bounded by $\varepsilon$.
The $\varepsilon>0$ is a predefined parameter before running the polyblock algorithm.
%, and can be set as any small positive number.
By following the polyblock algorithm, the detailed procedure for solving Problem \ref{prob:ts_upper_differential} is given as follows.
{\small \begin{algorithm}[H]
\caption{$\varepsilon$-optimal solution for Problem \ref{prob:ts_upper_differential}.}
\label{alg:ts_polyblock}
\begin{algorithmic}[1]
 \STATE
% Initial a point set $\mathcal{S}=\{s_1, s_2, ..., s_{|\mathcal{S}|}\}$, in which each point has two elements.
 Initialize a point set $\mathcal{Z}$ by a two-dimensional point $z_1 = (\alpha_{\max}, F(\alpha_{\max}))^T$, where the $\alpha_{\max}$ and $F(\alpha_{\max})$ indicate the maximal achievable value of the variable $\alpha$ and $z$, respectively.
% In other words, the two elements in point $s_1$ are $s_1(1)=T-\tau_{\min}$ and $s_1(2)=V(T)-V(\tau_{\min})$.
 \WHILE {$|\mathcal{Z}| >0$}
 \FOR {$l=1, 2, ...,|\mathcal{Z}|$}
 \STATE  Find the $u_l$ such that $ u_l z_l(1) F(u_l z_l(1))  + u_l z_l(2) = F(\alpha_{\max}) $ by utilizing the bisection search method, where $z_l(1)$ and $z_l(2)$ are the 1st and 2nd element of the vector $z_l$ respectively. Set $\Omega_l=u_l z_l$.
 \ENDFOR
 \STATE Find $l^*=\mathop {\arg \max} \limits_{1 \leq l \leq |\mathcal{Z}|} \left[F(\Omega_l(1))+\Omega_l(2) \right]$, where $\Omega_l(1)$ and $\Omega_l(2)$ are the 1st and 2nd element of the vector $\Omega_l$ respectively.
 \STATE For $\forall z \in \mathcal{Z}$, if there is $\left[F(z(1))+z(2) \right] \leq \left[F(\Omega_{l^*}(1))+\Omega_{i^*}(2) \right] +\varepsilon$, then delete the point $z$ from the set $\mathcal{Z}$.
 \IF {$|\mathcal{Z}| > 0 $}
 \STATE Search $j^* = \mathop {\arg \max} \limits_{1 \leq j \leq |\mathcal{Z}|} \left[F(z_j(1))+z_j(2) \right]$
 \STATE Find $u_{j^*}$ such that $u_{j^*} z_{j^*}(1) F(u_{j^*} z_{j^*}(1) ) + u_{j^*} z_{j^*}(2) = F(\alpha_{\max}) $ by utilizing the bisection search method. Set $\Omega_{j^*}=u_{j^*} z_{j^*}$.
 \STATE Generate two new points $z'=z_{j^*}+(\Omega_{j^*}-z_{j^*}) \circ (1,0)$ and $z''=z_{j^*}+(\Omega_{j^*}-z_{j^*}) \circ (0,1)$, where $\circ$ is the operation of Hadamard product.
 \STATE Add $z'$ and $z''$ into $\mathcal{Z}$, delete $z_{j^*}$ from the set $\mathcal{Z}$.
 \ENDIF
 \ENDWHILE
 \STATE Output the last $\Omega_{l^*}$ before $\mathcal{Z}$ is subtracted to be an empty set.
\end{algorithmic}
\end{algorithm} }

By following Algorithm \ref{alg:ts_polyblock}, the optimal solution of Problem \ref{prob:ts_upper_differential} can be achieved, which also paves the way for working out the optimal solution of Problem \ref{prob:ts_upper}.
To this end, Problem \ref{prob:ts_upper} can be solved optimally, which also indicates the optimal solving of Problem \ref{prob:ts}.

\section{Optimal Solution in PS Mode} \label{s:optimal_solution_PS}
In this section, Problem \ref{prob:ps} will be solved. It can be checked that Problem \ref{prob:ps} is a non-convex optimization problem either, considering the non-convexity of $\phi(x)$ under both logistic model and linear cut-off model in the objective function of Problem \ref{prob:ps}. In the following, we will show how to find the global optimal solution of
Problem \ref{prob:ps} under logistic model and linear cut-off model, respectively.
\subsection{The Case under Logistic Model} \label{s:optimal_solution_PS_nonlinear}
In this subsection, logistic model is adopted for the energy harvester, i.e., $\phi(x)$ is set to be the function in (\ref{e:model_nonlinear}).
To solve Problem \ref{prob:ps},
look into the objective function of Problem \ref{prob:ps},
the term $w_n \log \left(1 + \frac{p_n h_n (1-\beta_n)}{\sigma^2 w_n}\right)$ is monotonically decreasing function with $\beta_n$, and the term $w_n \log \left(1 + \frac{\phi(p_T h_n \beta_n)g_n}{\sigma^2 w_n}\right)$ is monotonically increasing function with $\beta_n$ considering the increasing monotonicity of the function $\phi(x)$.
Hence the maximal value of the term $w_n \log \left(1 + \frac{p_n h_n (1-\beta_n)}{\sigma^2 w_n}\right)$ and the term $w_n \log \left(1 + \frac{\phi(p_T h_n \beta_n)g_n}{\sigma^2 w_n}\right)$ is achieved when these two terms are equal, equivalently, there is
\begin{equation} \label{e:PS_two_term_equal}
p_n h_n (1-\beta_n) = \phi(p_T h_n \beta_n)g_n,
\end{equation}
which further indicates
\begin{equation} \label{e:PS_p_n_expression}
p_n = \frac{\phi(p_T h_n \beta_n)g_n}{h_n (1-\beta_n)}.
\end{equation}

Taking into account the fact that $p_n \leq p_T$ for $n\in \mathcal{N}$, and combine the constraint (\ref{e:PS_p_n_expression}), there is an implicit constraint
\begin{equation} \label{e:PS_nonlinear_power_implicit}
\frac{\phi(p_T h_n \beta_n)}{ (1-\beta_n)} \leq \frac{p_T h_n}{g_n}, \forall n \in \mathcal{N},
\end{equation}
which imposes an upper bound on $\beta_n$, denoted as $\beta_n^{U_{Y_1}}$, for $n\in \mathcal{N}$.
The $\beta_n^{U_{Y_1}}$ can be found by following bi-section search method such that
\begin{equation}
\frac{\phi(p_T h_n \beta_n^{U_{Y_1}})}{ (1-\beta_n^{U_{Y_1}})} = \frac{p_T h_n}{g_n}, \forall n \in \mathcal{N}.
\end{equation}
It can be easily derived that $\beta_n^{U_{Y_1}}<1$ since $\frac{p_T h_n}{g_n}$ is bounded for $n\in \mathcal{N}$.
Combining with the constraint (\ref{e:prob_ps_beta_max}), which indicates that $\beta_n \leq \frac{\phi^{-1}(q_{\max})}{p_T h_n}$,
and the constraint (\ref{e:prob_ps_beta_interval}), which indicates that $\beta_n \leq 1$, define $\beta_n^{U_Y} = \min(\beta_n^{U_{Y_1}}, \frac{\phi^{-1}(q_{\max})}{p_T h_n}, 1)$ for $n\in \mathcal{N}$, $\beta_n$ should satisfy
\begin{equation}
0 \leq \beta_n \leq \beta_n^{U_Y}, \forall n \in \mathcal{N}.
\end{equation}

Then by following the similar discussion for Problem \ref{prob:ts_lower_sim} in Section \ref{s:optimal_solution_TS}, solving Problem \ref{prob:ps} is equivalent with solving the following optimization problem
\begin{prob} \label{prob:ps_sim}
\begin{subequations}
\begin{align}
\max \limits_{\left\{\beta_n\right\}, \left\{w_n\right\}}~ & \sum \limits_{n=1}^N w_n \log \left(1 + \frac{\phi(p_T h_n \beta_n)g_n}{\sigma^2 w_n}\right)  \nonumber \\
\text{s.t.} \quad
& 0 \leq \beta_n \leq \beta_n^{U_Y}, \forall n \in \mathcal{N}, \\
%& p_n \geq 0, \forall n \in \mathcal{N}, \\
& w_n \geq 0, \forall n \in \mathcal{N}, \\
& \sum \limits_{n=1}^{N} w_n \leq w_T, \\
& \sum \limits_{n=1}^{N} \frac{\phi(p_T h_n \beta_n)g_n}{h_n (1-\beta_n)}  \leq p_T.
\end{align}
\end{subequations}
\end{prob}

By following the similar transformation from Problem \ref{prob:ts_lower_sim} to Problem \ref{prob:ts_lower_sim_no_w},
Problem \ref{prob:ps_sim} is equivalent with the following optimization problem
\begin{prob} \label{prob:ps_sim_no_w}
\begin{subequations}
\begin{align}
\max \limits_{\left\{\beta_n\right\}}~ & {\sum \limits_{n=1}^N \phi(p_T h_n \beta_n)g_n} \nonumber \\
\text{s.t.} \quad
& 0 \leq \beta_n \leq \beta_n^{U_Y}, \forall n \in \mathcal{N}, \\
& \sum \limits_{n=1}^{N} \frac{\phi(p_T h_n \beta_n)g_n}{h_n (1-\beta_n)}  \leq p_T \label{e:prob_ps_sim_no_w_cons}.
\end{align}
\end{subequations}
\end{prob}

Recalling $\phi(x)$ defined in (\ref{e:model_nonlinear}) is a monotonically increasing function,
thus both the objective function of Problem \ref{prob:ps_sim_no_w} and the left-hand side function of (\ref{e:prob_ps_sim_no_w_cons}) are increasing functions with the vector $\bm{\beta} \triangleq \left(\beta_1, \beta_2, ..., \beta_N\right)^T$.
Hence when $\sum \limits_{n=1}^{N} \frac{\phi(p_T h_n \beta_n^{U_Y})g_n}{h_n (1-\beta_n^{U_Y})}  \leq p_T$, the optimal solution is just set $\beta_n$ as large as possible, i.e., set $\beta_n = \beta_n^{U_Y}$ for $n\in \mathcal{N}$.
In general case, i.e., when $\sum \limits_{n=1}^{N} \frac{\phi(p_T h_n \beta_n^{U_Y})g_n}{h_n (1-\beta_n^{U_Y})}  >  p_T$,
Problem \ref{prob:ps_sim_no_w} also falls into the standard form of monotonic optimization problem.
Then by following the similar procedure in Algorithm \ref{alg:ts_polyblock}\footnote{Algorithm \ref{alg:ts_polyblock} works for a two-dimensional vector. The general solving algorithm can be found in \cite{Floudas:Opt} and is omitted due to the limit of space.}, the $\varepsilon$-optimal solution of Problem
\ref{prob:ps_sim_no_w} can be achieved.

In summary, the optimal solution of the original optimization problem in PS mode, i.e., Problem \ref{prob:ps}, can be achieved by following the steps in Algorithm \ref{alg:ps_summary}, i.e.,
\begin{algorithm}[H]
\caption{Searching procedure for the optimal solution of Problem \ref{prob:ps} under logistic model.}
\label{alg:ps_summary}
\begin{algorithmic}[1]
\IF {$\sum \limits_{n=1}^{N} \frac{\phi(p_T h_n \beta_n^{U_Y})g_n}{h_n (1-\beta_n^{U_Y})}  \leq p_T$ }
    \STATE {Set $\beta_n = \beta_n^{U_Y}$ for $n\in \mathcal{N}$.}
\ELSE
 \STATE {
 By following the similar procedure in Algorithm \ref{alg:ts_polyblock},
 find the optimal $\beta_n$ for $n\in \mathcal{N}$.
 }
  \ENDIF
 \STATE {
 Set $w_n = \frac{w_T \phi(p_T h_n \beta_n)g_n}{\sum_{n=1}^{N} \phi(p_T h_n \beta_n)g_n}$, where $\beta_n$ is calculated in Step 2 or Step 4 of Algorithm \ref{alg:ps_summary} for $n\in \mathcal{N}$.
  }
 \STATE{
 Calculate $p_n$ for $n \in \mathcal{N}$ according to (\ref{e:PS_p_n_expression}).
 }
 \STATE{
 Increase $p_n$ calculated in Step 6 to be $p'_{n}$ for $n\in \mathcal{N}$  such that $\sum_{n=1}^{N} p'_n = p_T$.
 }
 \STATE{
 Output $\beta_n$, $w_n$, and $p'_n$ for $n \in \mathcal{N}$.
 }
 \end{algorithmic}
\end{algorithm}

\subsection{The Case under Linear Cut-off Model} \label{s:optimal_solution_PS_cutoff}
In this subsection, linear cut-off model is adopted for the energy harvester, i.e., $\phi(x)$ is set to be the function in (\ref{e:model_linear_cutoff}).
Since the function $\phi(x)$ in (\ref{e:model_linear_cutoff}) is also a monotonically increasing function with $x$, by following Algorithm \ref{alg:ps_summary}, the optimal solution of Problem \ref{prob:ps} can be also achieved.
However, in this subsection, we will develop a simpler solution.

Looking into the expression of $\phi(x)$ in (\ref{e:model_linear_cutoff}), to guarantee positive energy harvested, there should be
\begin{equation} \label{e:ps_beta_positive_EH}
p_T h_n \beta_n \geq x_L, \forall n \in \mathcal{N},
\end{equation}
which indicates
\begin{equation} \label{e:ps_beta_lb}
\beta_n \geq \frac{x_L}{p_T h_n} \triangleq \beta_{n}^L, \forall n \in \mathcal{N}.
\end{equation}
On the other hand, when more than a power of $x_U$ is received at the energy harvester,
the energy harvester will become saturated. Thus there is no need to set the power of received energy to be larger than $x_U$, i.e., we have
\begin{equation}
  p_T h_n \beta_n \leq x_U, \forall n \in \mathcal{N},
\end{equation}
which implies
\begin{equation} \label{e:ps_beta_ub_0}
\beta_n \leq \frac{x_U}{p_T h_n}, \forall n \in \mathcal{N}.
\end{equation}

With the holding of constraints (\ref{e:ps_beta_lb}) and (\ref{e:ps_beta_ub_0}), there is
\begin{equation} \label{e:ps_phi_cutoff_sim}
\phi(x) = c(x-x_L).
\end{equation}
By following the similar discussion as in Section \ref{s:optimal_solution_PS_nonlinear},
we also have
\begin{equation} \label{e:PS_two_term_equal_cutoff}
p_n h_n (1-\beta_n) = \phi(p_T h_n \beta_n)g_n = c \left(p_T h_n \beta_n - x_L\right)g_n.
\end{equation}
which indicates
\begin{equation} \label{e:PS_p_n_expression_cutoff}
p_n = \frac{c\left(p_T h_n \beta_n - x_L\right)g_n}{h_n (1-\beta_n)}.
\end{equation}

Still following the similar discussion as in Section \ref{s:optimal_solution_PS_nonlinear},
the constraint (\ref{e:PS_nonlinear_power_implicit}) also holds for linear-cutoff model, which indicates
that $\beta_n$ is upper bounded by $\beta_n^{U_{Z_1}}$ ($\beta_n^{U_{Z_1}} < 1$), such that
\begin{equation}
\frac{c(p_T h_n \beta_n^{U_{Z_1}} - x_L)}{(1 - \beta_n^{U_{Z_1}})} = \frac{p_T h_n}{g_n}, \forall n \in \mathcal{N}.
\end{equation}
In addition, combine the constraint (\ref{e:prob_ps_beta_interval}) and constraint (\ref{e:prob_ps_beta_max}), \\
define $\beta_{n}^{U_Z} \triangleq \min\left(\beta_n^{U_{Z_1}}, \frac{\left(q_{\max} + c x_L\right)}{c p_T h_n}, \frac{x_U}{p_T h_n}, 1\right)$,
%constraint (\ref{e:ps_beta_ub_0}) can be enhanced to be
$\beta_n$ should be subject to the following constraint,
\begin{equation} \label{e:ps_beta_ub}
\beta_n \leq \beta_n^{U_Z}, \forall n \in \mathcal{N}.
\end{equation}

%Note that there is always $\beta_n^L < \beta_n^U$, i.e., $\frac{p_T h_{n}}{x_L} > 1$.
%If $\frac{p_T h_n}{x_L} \leq 1$, there will be no energy harvested on $n$th link, i.e., the link $n$ will be inactive.
%Without loss of generality, before investigating the Problem \ref{prob:ps}, the set $\{n'|\frac{p_T h_{n'}}{x_L} \leq 1 \}$ can be the excluded.
%Hence for $\forall n\in \mathcal{N}$, there is $\frac{p_T h_n}{x_L} >1$, i.e., $\beta_n^L < \beta_n^U$.

Then combining the constraints (\ref{e:ps_beta_lb}) and (\ref{e:ps_beta_ub}) and the expression of $\phi(x)$ in (\ref{e:ps_phi_cutoff_sim}), with the same discussion for the transformation from Problem \ref{prob:ps} to Problem \ref{prob:ps_sim}, to find the optimal solution of Problem \ref{prob:ps} under linear cut-off model for energy harvester, we only need to solve the following optimization problem
\begin{prob} \label{prob:ps_sim_cutoff}
\begin{subequations}
\begin{align}
\max \limits_{\left\{\beta_n\right\}, \left\{w_n\right\}}~ & \sum \limits_{n=1}^N w_n \log \left(1 + \frac{c\left(p_T h_n \beta_n - x_L\right)g_n}{\sigma^2 w_n}\right)  \nonumber \\
\text{s.t.} \quad
& \beta_n^L \leq \beta_n \leq \beta_n^{U_Z}, \forall n \in \mathcal{N}, \\
%& p_n \geq 0, \forall n \in \mathcal{N}, \\
& w_n \geq 0, \forall n \in \mathcal{N}, \\
& \sum \limits_{n=1}^{N} w_n \leq w_T, \\
& \sum \limits_{n=1}^{N} \frac{\phi(p_T h_n \beta_n)g_n}{h_n (1-\beta_n)}  \leq p_T,
\end{align}
\end{subequations}
\end{prob}
which can be simplified to be the following optimization problem by following the discussion method from Problem \ref{prob:ps_sim} to Problem \ref{prob:ps_sim_no_w}
\begin{prob} \label{prob:ps_sim_no_w_cutoff}
\begin{subequations}
\begin{align}
\max \limits_{\left\{\beta_n\right\}}~ & {\sum \limits_{n=1}^N c\left(p_T h_n \beta_n - x_L\right)g_n} \nonumber \\
\text{s.t.} \quad
& \beta_n^L \leq \beta_n \leq \beta_n^{U_Z}, \forall n \in \mathcal{N}, \\
& \sum \limits_{n=1}^{N} \frac{c\left(p_T h_n \beta_n - x_L\right)g_n}{h_n (1-\beta_n)}  \leq p_T \label{e:prob_ps_sim_no_w_cons_cutoff}.
\end{align}
\end{subequations}
\end{prob}

For Problem \ref{prob:ps_sim_no_w_cutoff}, it can be checked that the objective function of Problem \ref{prob:ps_sim_no_w_cutoff} is a linear function with $\beta_n$ for $n\in \mathcal{N}$, and the left-hand side function in (\ref{e:prob_ps_sim_no_w_cons_cutoff}) is convex with $\beta_n$ when $\frac{p_T h_n}{x_L} > 1$.
%\footnote{Note that if $\frac{p_T h_{n'}}{x_L} \leq 1$, there will be no energy harvested on $n'$th link, i.e., the link $n'$ will be inactive. Without loss of generality, before investigating the Problem \ref{prob:ps_sim_no_w_cutoff}, the set $\{n'|\frac{p_T h_{n'}}{x_L} \leq 1 \}$ can be the excluded.
%Hence for $\forall n\in \mathcal{N}$, there is $\frac{p_T h_n}{x_L} >1$.}
Therefore, Problem \ref{prob:ps_sim_no_w_cutoff} is a convex optimization problem.
It can be checked that Problem \ref{prob:ps_sim_no_w_cutoff} satisfies Slater's condition. Thus the KKT condition of Problem \ref{prob:ps_sim_no_w_cutoff} can serve as the sufficient and necessary condition of its optimal solution \cite{Boyd}, which can be written as
\begin{subequations}\label{e:KKT_cutoff}
\begin{align}
& c p_T h_n g_n - \Xi \frac{c g_n}{h_n} \frac{\left({p_T h_n} - x_L\right)}{\left(\beta_n -1 \right)^2}+ \Gamma_n - \Delta_n  = 0,
\label{e:ps_KKT_dbeta}  \\
%& \ln\left(1+\frac{p_n h_n}{w_n \sigma^2} \right)- \frac{p_n h_n}{p_n h_n+w_n \sigma^2}+ \mu_n - \nu = 0, \label{e:KKT_dw}  \\
& \Gamma_n \left(\beta_n - \beta_n^L \right)=0, \forall n \in \mathcal{N},  \label{e:ps_KKT_beta_lower}\\
& \Delta_n \left(\beta_n - \beta_n^{U_Z} \right)=0, \forall n \in \mathcal{N}, \label{e:ps_KKT_beta_upper} \\
& \Xi \left(p_T- \sum \limits_{n=1}^{N} \frac{c\left(p_T h_n \beta_n - x_L\right)g_n}{h_n (1-\beta_n)} \right) =0, \label{e:ps_KKT_beta_sum}\\
& \Gamma_n \geq 0, \Delta_n \geq 0,  \forall n \in \mathcal{N}, \label{e:ps_KKT_positive_lagrange_n}\\
& \Xi \geq 0, \label{e:ps_KKT_positive_lagrange} \\
& \text{Constraints} (\ref{e:ps_beta_lb}), (\ref{e:ps_beta_ub}), (\ref{e:prob_ps_sim_no_w_cons_cutoff}).
\end{align}
\end{subequations}

According to (\ref{e:ps_KKT_beta_lower}) and (\ref{e:ps_KKT_beta_upper}), when $\beta_n > \beta_n^L$ and $\beta_n < \beta_n^{U_Z}$, there is $\Gamma_n = 0$ and $\Delta_n$, respectively.
So when $\beta_n^L < \beta_n < \beta_n^{U_Z}$, $\Gamma_n =\Delta_n =0$,
%Define the set $\mathcal{B} = \{n|\frac{x_L}{p_T h_n} < \beta_n < \frac{x_U}{p_T h_n}, n \in \mathcal{N} \}$, and
there is
\begin{equation} \label{e:ps_beta_Xi}
\beta_n = 1 - \sqrt{\frac{\Xi\left(p_T h_n - x_L \right)}{p_T h_n^2}}
\end{equation}
according to (\ref{e:ps_KKT_dbeta}).

For a given $\Xi$, if the calculated $\beta_n$ by following (\ref{e:ps_beta_Xi}) is larger than its upper bound $\beta_n^{U_Z}$, then $\beta_n = \beta_n^{U_Z}$ by checking (\ref{e:ps_beta_ub}), (\ref{e:ps_KKT_dbeta}) and (\ref{e:ps_KKT_beta_upper}).
Similarly, if the calculated $\beta_n$ by following (\ref{e:ps_beta_Xi}) is smaller than its lower bound $\beta_n^L$,
then $\beta_n = \beta_n^L$.
Hence $\beta_n$ can be expressed by $\Xi$ in a precise way as follows
%\begin{equation} \label{e:ps_beta_Xi_complete}
%\beta_n(\Xi)= \left[1 - \sqrt{\frac{\Xi\left(p_T h_n - x_L \right)}{p_T h_n^2}}\right]\Bigg|_{\frac{x_L}{p_T h_n}}^{\frac{x_U}{p_T h_n}}, \forall n\in \mathcal{N},
%\end{equation}
\begin{equation} \label{e:ps_beta_Xi_complete}
\beta_n(\Xi)= \left[1 - \sqrt{\frac{\Xi\left(p_T h_n - x_L \right)}{p_T h_n^2}}\right]\Bigg|_{\beta_n^L}^{\beta_n^{U_Z}}, \forall n\in \mathcal{N},
\end{equation}
where the operation $[x]|_a^b = \max(a, \min(x,b))$.
The $\beta_n(\Xi)$ is actually a monotonically decreasing function with $\Xi$ for $n\in \mathcal{N}$.

On the other hand, it should be noticed both the left-hand side function of (\ref{e:prob_ps_sim_no_w_cons_cutoff}) and the objective function of Problem \ref{prob:ps_sim_no_w} are increasing functions with $\beta_n$ for $n\in \mathcal{N}$, so it is better to set $\beta_n$ as large as possible, which indicates that the optimal solution of Problem \ref{prob:ps_sim_no_w_cutoff} happens when the constraint (\ref{e:prob_ps_sim_no_w_cons_cutoff}) become active, i.e.,
\begin{equation} \label{e:prob_ps_sim_no_w_cons_cutoff_active}
\sum \limits_{n=1}^{N} \frac{c\left(p_T h_n \beta_n(\Xi) - x_L\right)g_n}{h_n (1-\beta_n(\Xi))}  = p_T.
\end{equation}
Given that the left-hand side function of constraint (\ref{e:prob_ps_sim_no_w_cons_cutoff_active}) is increasing with $\beta_n$, and the monotonicity of $\beta_n(\Xi)$ with $\Xi$.
The left-hand side function of constraint (\ref{e:prob_ps_sim_no_w_cons_cutoff_active}) is also monotonic with $\Xi$.
Hence the $\Xi$ such that the equality (\ref{e:prob_ps_sim_no_w_cons_cutoff_active}) holds can be searched by following bi-section method.
Note that when
\begin{equation} \label{e:prob_ps_sim_no_w_cons_check_term}
\sum \limits_{n=1}^{N} \frac{c\left(p_T h_n \beta_n(0) - x_L\right)g_n}{h_n (1-\beta_n(0))},
\end{equation}
which is the maximal value of the left-hand side of (\ref{e:prob_ps_sim_no_w_cons_cutoff_active}) is less than $p_T$, the optimal configuration of $\beta_n$ is $\beta_n = \beta_n^{U_Z}$ for $n\in \mathcal{N}$.

In the end of this subsection, the simple solution for PS mode under linear cut-off model is summarized as follows
\begin{algorithm}[H]
\caption{Searching procedure for the optimal solution of Problem \ref{prob:ps} under linear cut-off model.}
\label{alg:ps_summary_cutoff}
\begin{algorithmic}[1]
  %\STATE {Check the whether the term in .}
  %\IF {The term in (\ref{e:prob_ps_sim_no_w_cons_check_term}) is less than $p_T$}
  \IF {$\sum \limits_{n=1}^{N} \frac{c\left(p_T h_n \beta_n(0) - x_L\right)g_n}{h_n (1-\beta_n(0))} \leq p_T$}
    \STATE {Set $\beta_n = \beta_n^{U_Z}$ for $n\in \mathcal{N}$.}
  \ELSE
  \STATE {
 Use bi-section search method to find the $\Xi$ such that equality (\ref{e:prob_ps_sim_no_w_cons_cutoff_active}) holds.
     }
 \STATE {
 Calculate $\beta_n(\Xi)$ according to (\ref{e:ps_beta_Xi_complete}) for $n\in \mathcal{N}$.
  }
  \ENDIF
 \STATE{
 Calculate $p_n$ for $n \in \mathcal{N}$ according to (\ref{e:PS_p_n_expression_cutoff}).
 }
 \STATE{
 Set $w_n = \frac{w_T c\left(p_T h_n \beta_n - x_L\right)g_n}{\sum_{n=1}^{N} c\left(p_T h_n \beta_n - x_L\right)g_n}$, where $\beta_n$ is calculated in Step 2 or Step 5 of Algorithm \ref{alg:ps_summary_cutoff} for $n\in \mathcal{N}$.
 }
 \STATE{
 Increase $p_n$ calculated in Step 3 to be $p'_{n}$ for $n\in \mathcal{N}$  such that $\sum_{n=1}^{N} p'_n = p_T$.
 }
 \STATE{
 Output $\beta_n$, $w_n$, and $p'_n$ for $n \in \mathcal{N}$.
 }
 \end{algorithmic}
\end{algorithm}

\section{Numerical Results} \label{s:num}
In this section, numerical results are presented to verify the effectiveness of our proposed methods.
The system parameters are set as follows in default.
There are 4 relay nodes, i.e., $N=4$ and $\mathcal{N} = \{1, 2, 3, 4\}$. The total system bandwidth $w_T = 1$ MHz. The total transmit power of source node $p_T = 1$ W.
The power spectrum density of noise $\sigma^2 = 1\times 10^{-14}$ W/Hz.
The maximal transmit power of every relay node $q_{\max} = 50$ mW.
The carrier frequency is set as 1 GHz.
%Both $g_n$ and $h_n$ for $n \in \mathcal{N}$ are randomly distributed between -38.47dB and -52.45dB,
Both $g_n$ and $h_n$ for $n \in \mathcal{N}$ are uniformly distributed between -40dB and -50dB,
which approximately correspond to the attenuation in free space between 2m and 10m, respectively.
For the energy harvester, by utilizing the curve fitting tool on the measured data points in Fig.\ref{fig:nonlinear}, it is calculated that $M=2.3 \times 10^{-2}$, $a=170$, and $b=1.398 \times 10^{-2}$ for logistic model, and $c=0.7833$, $x_L=0$, and $x_U = 3\times 10^{-2}$ for linear cut-off model.
When running the polyblock algorithm, $\varepsilon$ is set as $1\times 10^{-2}$.
As a comparison, a relay selection method, which usually appears in literatures \cite{TWC_2014_Poor}, is implemented.
In the relay selection method, all the allowable transmit power and bandwidth are imposed on one link from the source node through some relay node to the destination node. The link associated with the maximal end-to-end throughput is selected.
With the implementation of relay selection method, there are also two modes: TS mode and PS mode. For the ease of presentation, we will denote the relay selection method under TS mode and PS mode as ``TS-select'' and ``PS-select'', respectively.

\begin{figure*}
\centering
\subfloat[The case under logistic model.]{
        \label{f:logistic_p_T}
        \includegraphics[width=1.0 \figwidth]{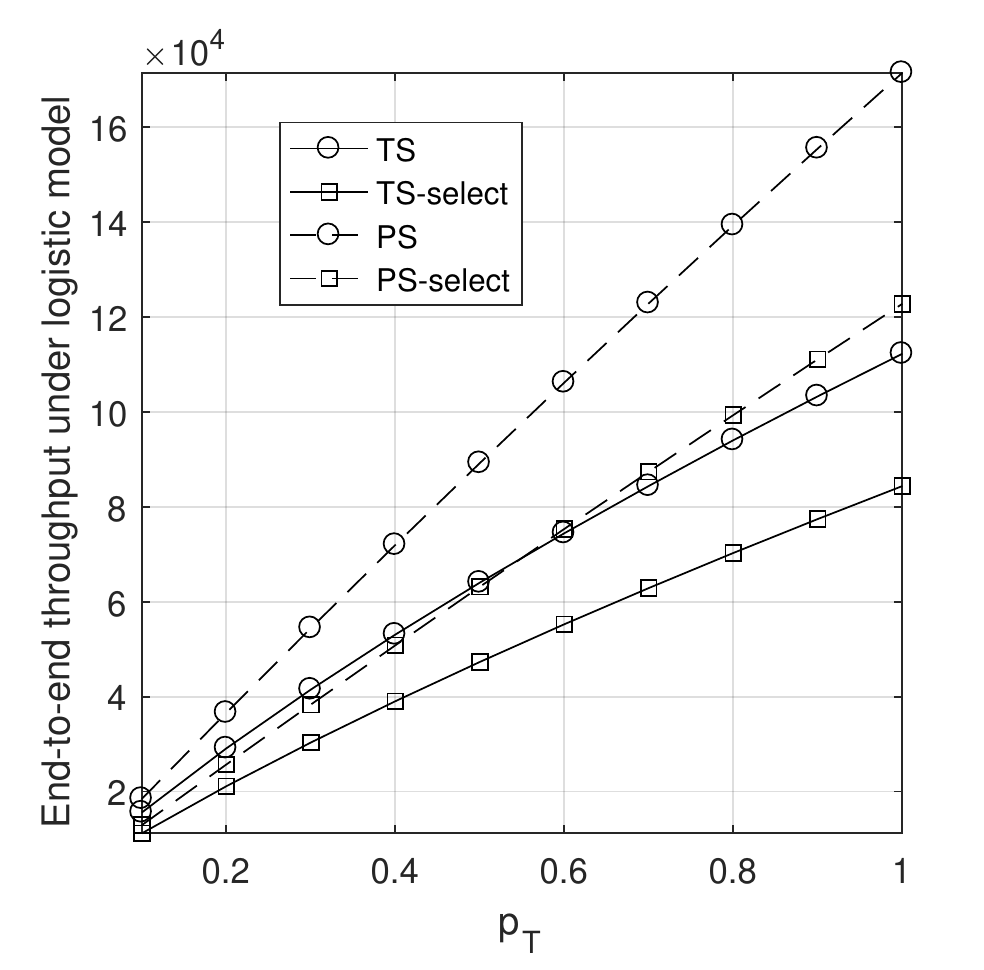}}
\subfloat[The case under linear cut-off model.]{
        \label{f:linear_p_T}
        \includegraphics[width=1.0 \figwidth]{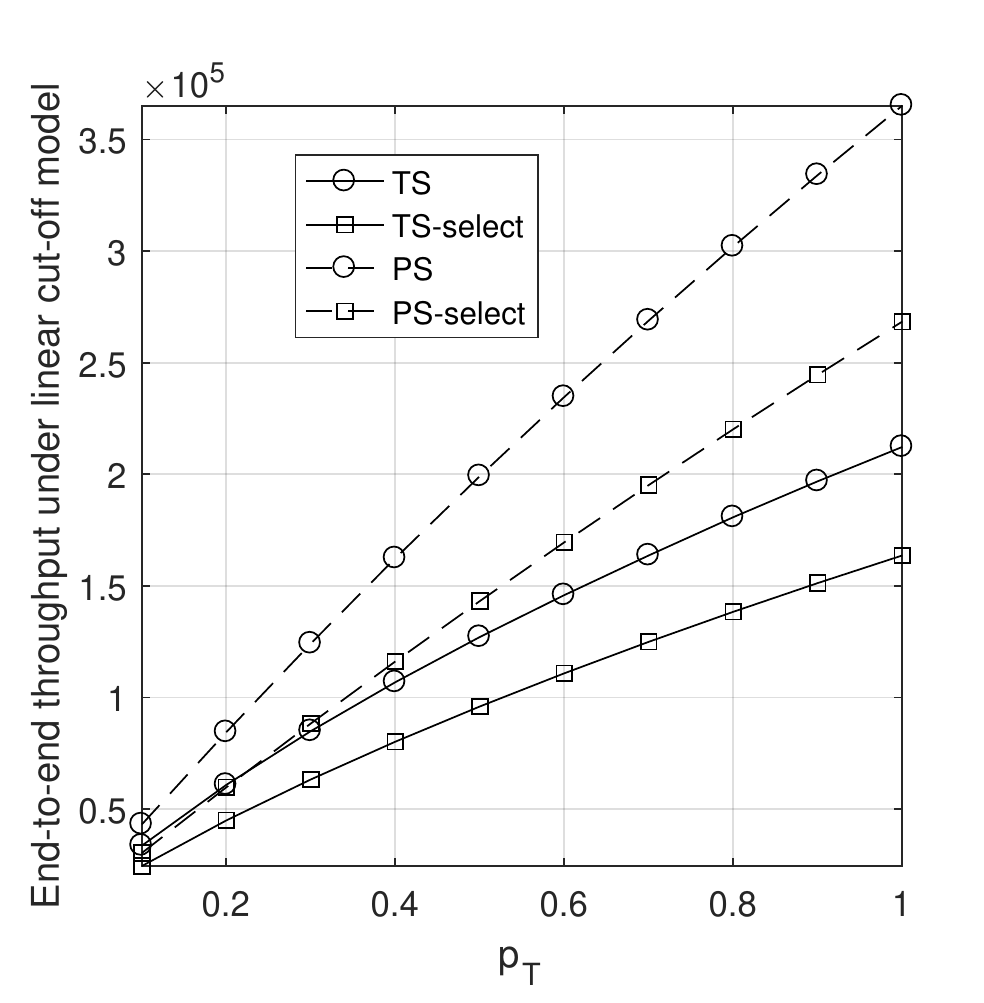}}
				
\caption{End-to-end throughput versus transmit power $p_T$.}
\label{f:throughput_vs_p}
\end{figure*}

In Fig. \ref{f:throughput_vs_p}, the end-to-end throughput from the source node to the destination node is plotted versus the transmit power $p_T$ for TS mode, TS-select mode, PS mode, and PS-select mode respectively, under logistic model (in Fig. \ref{f:logistic_p_T}) and under linear cut-off model (in Fig. \ref{f:linear_p_T}).
It can be observed that as $p_T$ grows, the end-to-end throughput under every mode grows, which is in coordination with intuition.
Additionally, it can be seen that PS mode outperforms PS-select mode and TS mode outperforms TS-select mode, which verifies the effectiveness of our proposed method.
Moreover, it can be also observed that PS mode always outperforms TS mode. This is consistent with the results in existing literatures \cite{Zhang_TC_2013,Saman_Access} on SWIPT and provides helpful suggestion for the implementation in real application.
% THE REASON FOR THE DIFFERENCE BETWEEN LOGISTIC AND LINEAR CUT-OFF IS BECAUSE THE RECEIVED POWER IS AT THE SCALE 1E-4 AND 1E-5, IN WHICH REGION, THE LINEAR CUTOFF MODEL GENERATE A TWICE OUTPUT COMPARED TO LOGISTIC MODEL.
%Last but not least, the performance under logistic model and linear cut-off model are nearly the same. This comes from the nearly identical approximation effect on the feature of energy harvester between logistic model and linear cut-off model.
%Given the nearly same performance, linear cut-off model will be preferred in PS mode in practise, as less computation complexity will be involved compared with the polyblock algorithm for logistic model.

\begin{figure*}
\centering
\subfloat[The case under logistic model.]{
        \label{f:logistic_w_T}
        \includegraphics[width=1.0\figwidth]{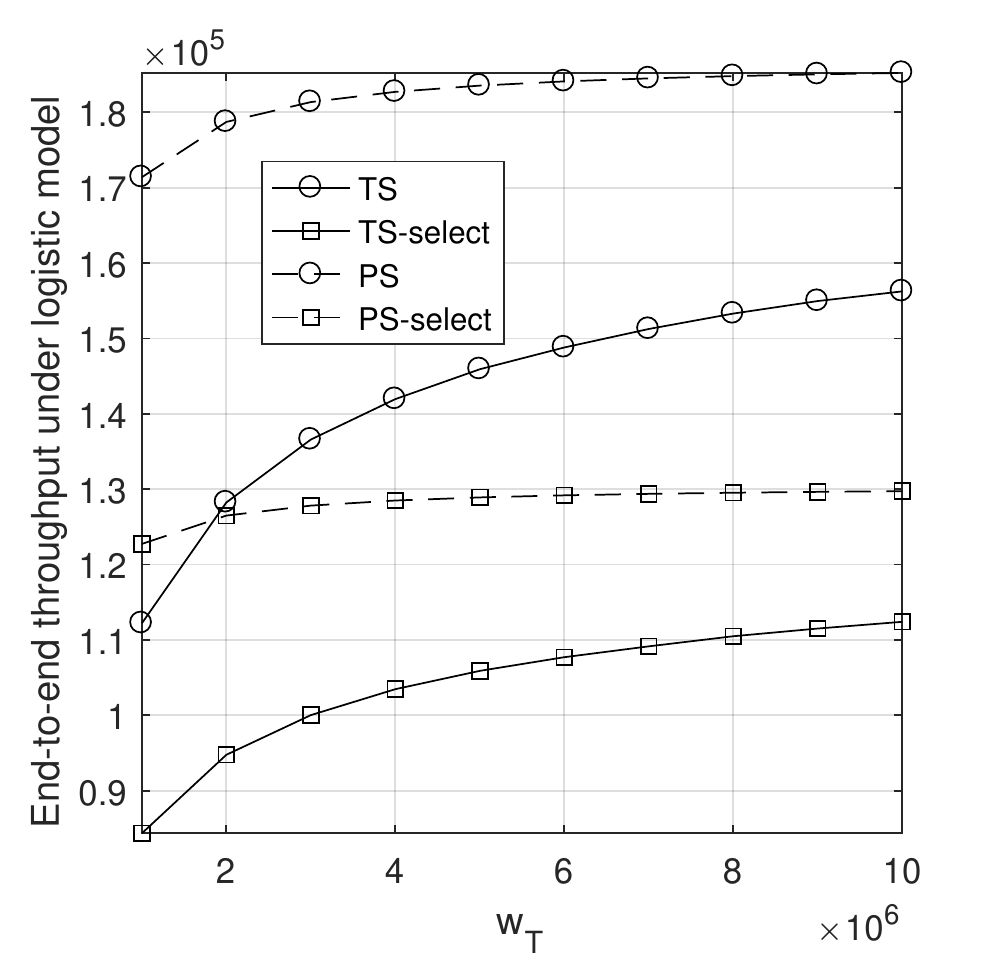}}
\subfloat[The case under linear cut-off model.]{
        \label{f:linear_w_T}
        \includegraphics[width=1.0\figwidth]{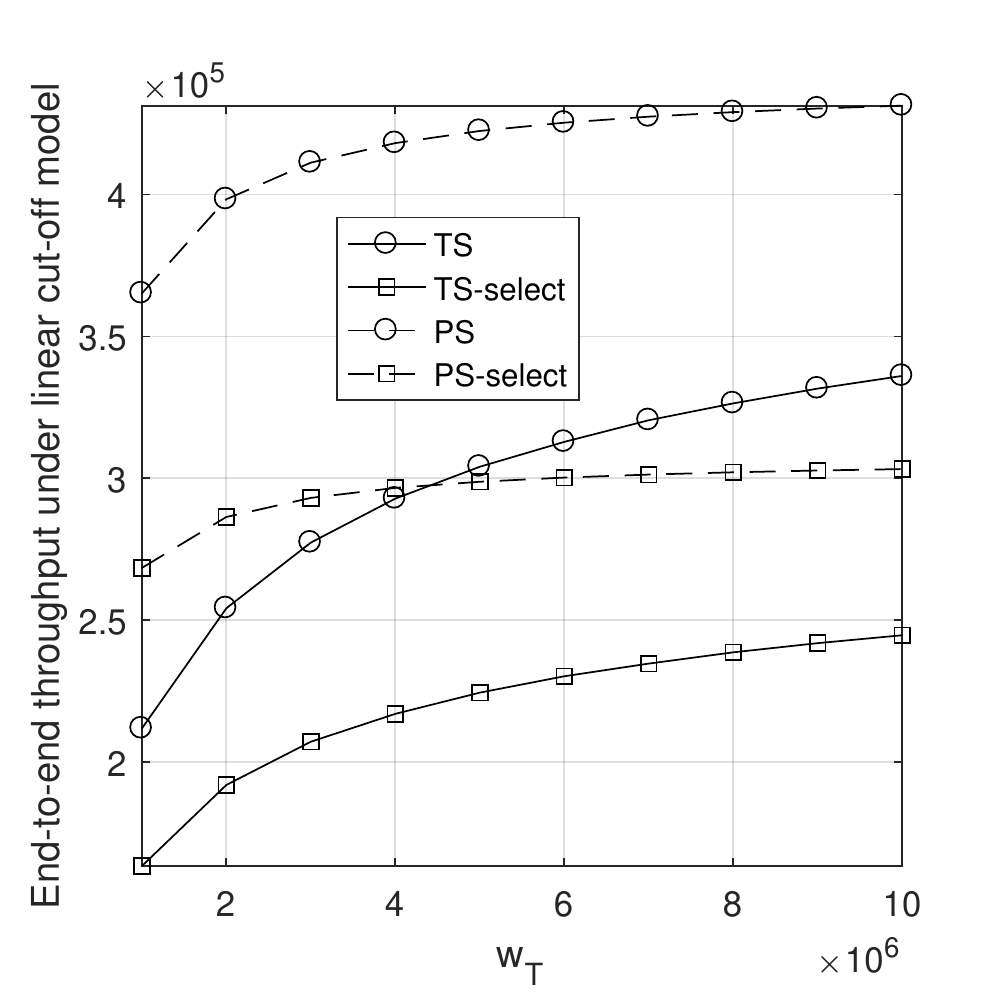}}
				
\caption{End-to-end throughput versus system bandwidth $w_T$.}
\label{f:throughput_vs_w}
\end{figure*}

In Fig. \ref{f:throughput_vs_w}, the end-to-end throughput is plotted versus system bandwidth $w_T$ for TS mode, TS-select mode, PS mode, and PS-select mode respectively, under logistic model (in Fig. \ref{f:logistic_w_T}) and under linear cut-off model (in Fig. \ref{f:linear_w_T}).
Similar observations can be obtained as for Fig. \ref{f:throughput_vs_p}.
The only difference from Fig. \ref{f:throughput_vs_p} lies in that the end-to-end throughput grows with $w_T$ at a decreasing rate, rather than a nearly constant rate.
This indicates that increasing total transmit power $p_T$ will play a more significant effect on improving end-to-end throughput compared with increasing the system bandwidth $w_T$.

\begin{figure*}
\centering
\subfloat[The case under logistic model.]{
        \label{f:logistic_SNR}
        \includegraphics[width=1.0 \figwidth]{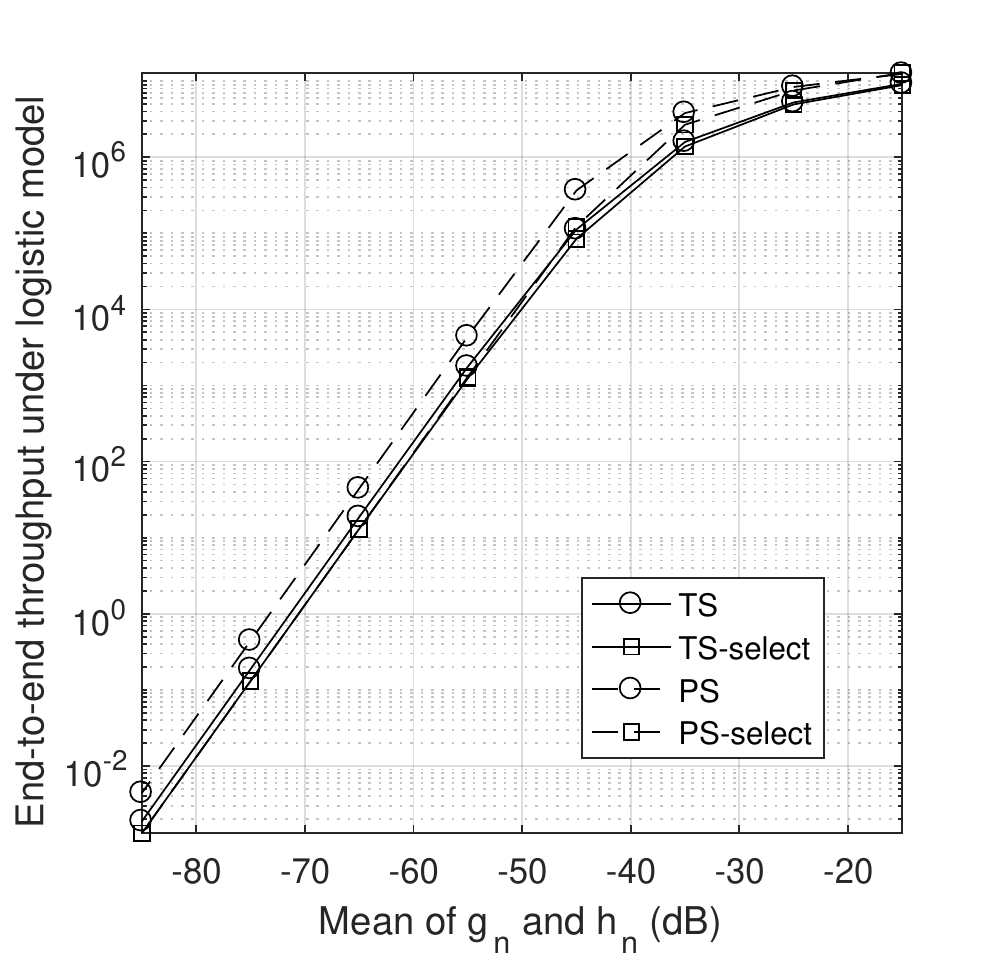}}
\subfloat[The case under linear cut-off model.]{
        \label{f:linear_SNR}
        \includegraphics[width=1.0 \figwidth]{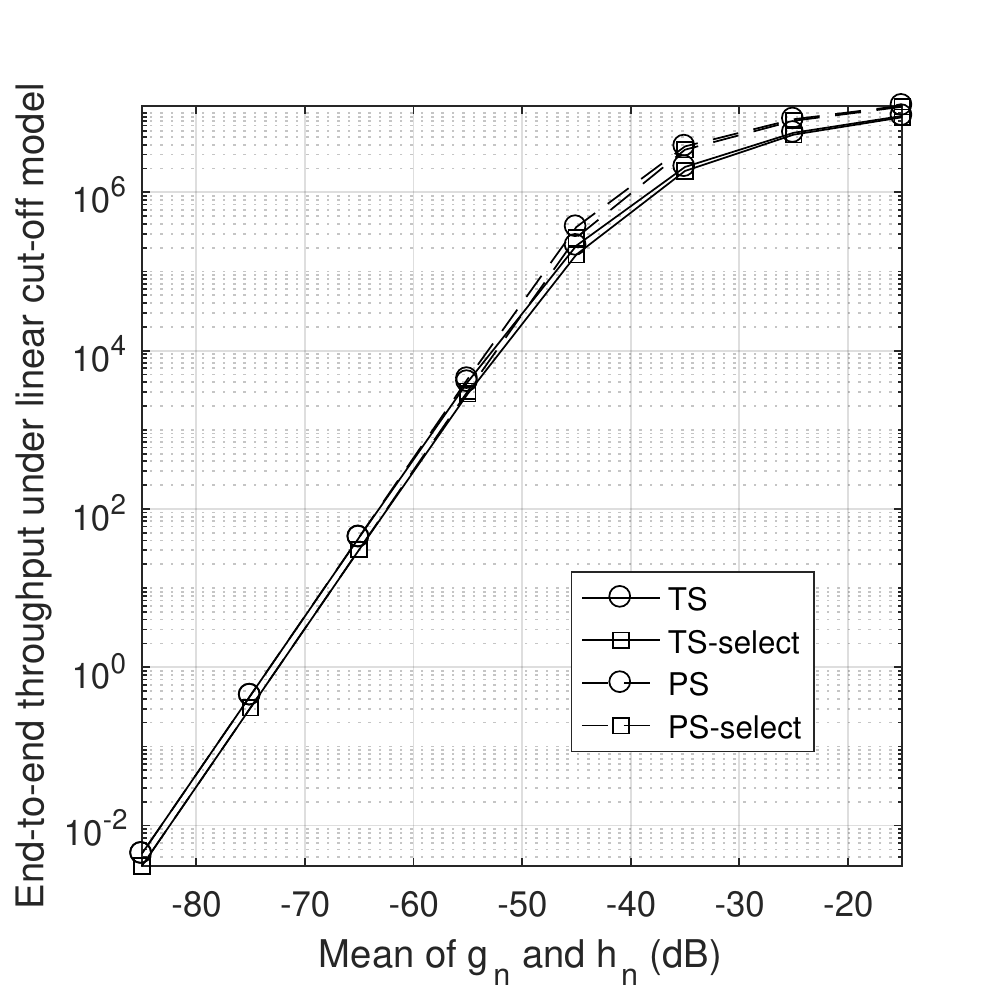}}				
\caption{End-to-end throughput versus channel gain $g_n$ and $h_n$.}
\label{f:throughput_vs_SNR}
\end{figure*}

In Fig. \ref{f:throughput_vs_SNR}, the end-to-end throughput is plotted versus the mean of $g_n$ and $h_n$ for TS mode, TS-select mode, PS mode, and PS-select mode respectively, under logistic model (in Fig. \ref{f:logistic_SNR}) and under linear cut-off model (in Fig. \ref{f:linear_SNR}).
Note that when the mean of $g_n$ and $h_n$ are set as $x$ dB.
Then the $g_n$ and $h_n$ are uniformly distributed between $[x-5, x+5]$ dB.
Similar observations can be obtained as for Fig. \ref{f:throughput_vs_p} as well.
It can be also seen that the value of $g_n$ and $h_n$ have great influence on the end-to-end throughput.
This indicates such a suggestion: We should try the best to place relay nodes at the locations close to source node and destination node with little shadowing and fading.

\section{Conclusion} \label{s:conclusion}
In this paper, end-to-end throughput is maximized for a two-hop DF multiple-relay network implemented with SWIPT under TS mode and PS mode. Transmit power and bandwidth on every link from source to destination, and the PS ratio or TS ratio on every relay node are optimized. Two types of nonlinear model are adopted for the energy harvester. For every combinational case in terms of working mode and nonlinear model, an optimization problem is formulated, all of which are non-convex. With a series of analysis and transformation, and with the aid of bi-level optimization and monotonic optimization, etc., we find the global optimal solution for the optimization problem in every case. In some case, the offered optimal solution is closed-form or semi-closed-form.
Our findings can provide helpful suggestion for the application of SWIPT-powered relay network in the future.

%This paper investigates the end-to-end throughput maximization problem for a two-hop multiple-relay network, with relays powered by simultaneous wireless information and power transfer (SWIPT) technique. Nonlinearity of energy harvester at every relay node is taken into account and two existing approximation models are adopted: logistic model and linear cut-off model. Decode-and-forward (DF) is implemented. Power splitting (PS), time switching (TS), and hybrid mode which is a combination of PS mode and TS mode are considered, respectively. Optimization problem are formulated under PS mode, TS mode, and hybrid mode under logistic model and linear cut-off model. End-to-end throughput is aimed to be maximized by optimizing the transmit power and bandwidth on every source-relay-destination link, and PS ratio and/or TS ratio on every relay node. Although the formulated optimization problems are all non-convex. Through a series of analysis and transformation, and with the aid of bi-level optimization and monotonic optimization, etc.,  we find the global optimal solution for the optimization problem in every case. In some cases, simple yet optimal solution is also derived.
%Numerical results verifies the effectiveness of our proposed methods.

%\appendices
%\section{Convexity Proof of Problem P2}\label{a:convexity_P2}

%\section*{Acknowledgments}

\end{document}